\documentclass{aastex631}
\usepackage{graphicx}
\usepackage{natbib}
\usepackage{float}
\usepackage{colortbl}
\usepackage[caption=false]{subfig}
\usepackage{wrapfig}
\usepackage{amsmath}
\usepackage{latexsym}
\usepackage{amssymb}
\usepackage{longtable}
\usepackage{epsf}

\graphicspath{{Figures/}}
\definecolor{QGray}{gray}{0.6}

\bibpunct{(}{)}{;}{a}{}{,}

\begin{document}

\title{Plausible constraints on the range of bulk terrestrial exoplanet compositions in the Solar neighbourhood}

\author{Rob J. Spaargaren}
\affiliation{Institute of Geophysics, ETH Zurich\\
Sonneggstrasse 5\\
8092 Zurich, Switzerland}

\author{Haiyang S. Wang}
\affiliation{Institute for Particle Physics and Astrophysics, ETH Zurich\\
Wolfgang-Pauli-Strasse 27\\
8093 Zürich, Switzerland}

\author{Stephen J. Mojzsis}
\affiliation{Origins Research Institute, Research Centre for Astronomy and Earth Sciences, MTA Centre of Excellence \\
15-17 Konkoly Thege Miklos ut\\
Budapest, 1121 Hungary}
\affiliation{Department of Lithospheric Research, University of Vienna\\
UZA 2, Josef-Holaubek-Platz 2\\
A-1090 Vienna, Austria}
\affiliation{Institute for Earth Sciences, Friedrich-Schiller University\\
Burgweg 11\\
07749 Jena, Germany}
\affiliation{Department of Geological Sciences, University of Colorado
UCB 399,2200 Colorado Avenue\\
Boulder, 80309 Colorado, USA}

\author{Maxim D. Ballmer}
\affiliation{Department of Earth Sciences, University College London\\
Gower Place\\
London WC1E 6BT, United Kingdom}

\author{Paul J. Tackley }
\affiliation{Institute of Geophysics, ETH Zurich\\
Sonneggstrasse 5\\
8092 Zurich, Switzerland}


\date{Received ... / Accepted ...} 
\begin{abstract}
Rocky planet compositions regulate planetary evolution by affecting core sizes, mantle properties, and melting behaviours. Yet, quantitative treatments of this aspect of exoplanet studies remain generally under-explored. 
We attempt to constrain the range of potential bulk terrestrial exoplanet compositions in the solar neighbourhood ($<$200 pc).
We circumscribe probable rocky exoplanet compositions based on a population analysis of stellar chemical abundances from the Hypatia and GALAH catalogues. 
We apply a devolatilization model to simulate compositions of hypothetical, terrestrial-type exoplanets in the habitable zones around Sun-like stars, considering elements O, S, Na, Si, Mg, Fe, Ni, Ca, and Al. We further apply core-mantle differentiation by assuming constant oxygen fugacity, and model the consequent mantle mineralogy with a Gibbs energy minimisation algorithm. 
We report statistics on several compositional parameters and propose a reference set of (21) representative planet compositions for using as end-member compositions in imminent modelling and experimental studies. 
We find a strong correlation between stellar Fe/Mg and metallic core sizes, which can vary from 18 to 35 wt\%. 
Furthermore, stellar Mg/Si gives a first-order indication of mantle mineralogy, with high-Mg/Si stars leading to weaker, ferropericlase-rich mantles, and low-Mg/Si stars leading to mechanically stronger mantles. The element Na, which modulates crustal buoyancy and mantle clinopyroxene fraction, is affected by devolatilization the most.
While we find that planetary mantles mostly consist of Fe/Mg-silicates, core sizes and relative abundances of common minerals can nevertheless vary significantly among exoplanets. These differences likely lead to different evolutionary pathways among rocky exoplanets in the solar neighbourhood.
\end{abstract}


\section{Introduction} \label{sec:Intro}
A plethora of rocky exoplanets has been documented around stars in the solar neighbourhood \citep{ps}. Based on formation models, it is expected that a solar-type star (i.e., an F, G, or K-type star) 
will host at least one planet \citep{mulders2018,Bryson2020}. Thus, we can expect the catalogue of discovered rocky exoplanets to grow substantially in the future. Based on first-order interpretations of mass-radius-density relationships, these ``terrestrial-type" planets are inferred to share many similarities with the rocky planets in our Solar System, e.g.\ in terms of a general, layered structure with a metallic iron core, a silicate mantle and crust, and a relatively low-mass atmosphere compared to giant planets. Those worlds not represented in our Solar System, the so-called Super-Earths and sub-Neptunes, are not treated here. These three fundamental layers - modulated by the physical, chemical and mechanical properties of their constituent materials - will differ in both mass and geophysical expression(s) between planets. Consequently, rocky exoplanets may be expected to follow different evolutionary pathways and thus sustain surface conditions that deviate, perhaps markedly, from Earth, Venus or Mars. Owing to the fact that our current understanding of rocky planets is based on what we know from studying Solar System planets, we must be prepared to challenge our assumptions of what a terrestrial-type planet is and what range of properties it could plausibly assume. As we continue to expand our methods of studying exoplanets, it makes sense to consider credible and testable physical-chemical bounds for rocky exoplanets that can help guide remote observations \citep{Mojzsis2022}.

An important consideration that arises in terrestrial-type exoplanet studies is whether such planets are capable of supporting mobile crust (i.e. plate tectonics-like behaviour). Alternatively, silicate+metal planets can be locked in a geodynamical regime with very little or no surface mobility (stagnant lid), amongst other possibilities \citep[e.g., the heat pipe regime;][]{Moore2013,Lourenco2020}. Whether a planet can sustain effective crustal recycling by processes such as plate tectonics is considered an important factor in determining its potential to host biological activity over geologic timescales \citep[e.g.,][]{Parnell2004,Noack2013,Mojzsis2021}. Evidently, planet size affects thermal evolution and propensity towards plate tectonics \citep[e.g., ][]{Valencia2006,ONeill2007,vanHeck2011,Stamenkovic2012,Stein2013}. Furthermore, interior planet structure and mantle rheology modulate thermal evolution and hence propensity toward plate tectonics \citep{Noack2014,Stamenkovic2016,Guerrero2018}. The latter properties are modulated by bulk composition, which is a parameter that has been hitherto less studied in the exoplanetary context \citep{Shahar2019}.

While the vast majority of rocky planets are thought to consist of a metallic iron core, a rocky mantle consisting of mainly Si, Mg, and O, and a volatile atmosphere, the relative abundances of these elements will have far-reaching effects on the planet's interior properties and evolutionary pathway. The eventual size of the metallic core is a direct function of ambient Fe abundance and oxygen fugacity of the planetary source material \citep{Corgne2008,Rubie2015}. The major-element composition of the silicate reservoir of a planet defines its mantle mineralogy, which in turn controls physical mantle parameters, such as density and viscosity \citep{Takeda1998,Yamazaki2001}, as well as melting behaviour \citep{Hirschmann2000,Kiefer2015}. Moreover, bulk planetary composition ultimately affects the atmospheric evolution through altering interaction between the interior and the atmosphere \citep[e.g.,][]{Spaargaren2020}. The interior of an exoplanet can only be decoded by analysis of its atmosphere, but we need to know how interior composition affects atmospheric evolution to make sense of atmospheric observations. This work helps build the geochemical foundations for this goal, by establishing
the plausible range of compositions that rocky planets can attain from analysis of stellar abundances within 200 pc of the Sun.

\subsection{Star-planet compositional link}
Rocky exoplanet composition cannot be directly observed, but it can be constrained by considering that planet composition is linked to that of its host star. A rocky planet forms from condensing material in the planet-forming disc, which consists of the same material as the forming star. Therefore, planet composition can, in principle, be estimated using the host star composition by considering compositional fractionation during the planet formation process. Recent evidence from polluted white dwarf stars, which are stars actively accreting (exo)planetary material, demonstrates that planetary compositions indeed largely mirror stellar abundances \citep{Doyle2019,Bonsor2021}.

While the compositional link between a planet and its host star is intrinsic, it behaves differently for elements with different volatilities \citep{Halliday2001, Sossi2019, Wang2019}. Comparisons made between Earth and Sun show that their compositions are very similar for elements with high condensation temperature in the planet-forming disc \citep[refractory elements, such as the rare earth elements, Al and Ca; ][]{Lodders2003}, and Earth follows a depletion trend in volatile elements; i.e., elements with lower condensation temperature, such as Na, K, Rb, and Cs \citep[e.g.][]{Halliday2001,Palme2013, Wang2019}. Similar observations have also been found towards other solar system rocky bodies (particularly Mars and a variety of chondrites) relative to the Sun \citep{Bland2005, Sossi2018, Yoshizaki2020}. This solar-system-based observational phenomenon is dubbed as ``devolatilization" and hypothesized to be a universal process in the formation of rocky planets \citep{Wang2019}. In \cite{Doyle2019}, it is found that the oxidation state of extrasolar rocks inferred from the abundance measurements of polluted white dwarf atmospheres are overall consistent with that of the solar system rocky bodies, suggesting that a similar refractory-volatile fractionation process might have also happened in the early exoplanet systems. Indeed, in \cite{Harrison2021}, it is evident that post-nebular devolatilization of moderately volatiles (e.g. Na) is present in rocky exoplanetary materials that polluted white dwarfs. This offers an importance piece of observational evidence for the aforementioned hypothesis. 

A typical trend of devolatilization, as firstly quantified in \cite{Wang2019} based on the bulk Earth and proto-Sun, reflects the observed behaviours for refractories and for (moderately) volatiles as described above. The yet-limited bulk compositional data for other solar system terrestrial planets still inhabit an adequate quantitative model to be made \citep{Lin2022}, but it has been suggested in \cite{Wang2019MNRAS, Wang2022_planethoststars} that such a trend for Mars and Venus may not be significantly different from that of Earth by taking into account the large uncertainties in their individual \textit{bulk} compositions \citep[e.g.][]{Morgan1980, Wang2018_Earthcomp, Yoshizaki2020}. As a first-order application for a population analysis as aimed here, we choose to adopt the well quantified Sun-to-Earth devolatilization trend \citep{Wang2019} to take into account this important observational effect while studying rocky exoplanet bulk compositions from the measured host stellar chemical compositions (Sects. \ref{ssec:Metho_stars} \& \ref{ssec:Metho_star2pl}). This will allow us to study the effects of both the intrinsic spread of the stellar chemical compositions as well as the reduced concentrations of volatile components on the geochemical and geophysical properties of rocky exoplanets, particularly those in the habitable zone around Sun-like stars. More discussion on the applicability and limitation of the approach can be found in Sect. \ref{ssec:Disc_devol}.



Although stellar abundances have been proposed for constraining rocky exoplanet compositions, the foci remain on the refractory elements \citep[e.g., Mg, Si, and Fe; ][]{Dorn2015, Unterborn2016,Dorn2017, Putirka2019} or on the individual planets \citep{Wang2019MNRAS,Wang2022_alphacentauri, Wang2022_planethoststars}. Population analysis of stellar abundances - while taking into account the devolatilization effect for estimating rocky exoplanet compositions - is a gap in the literature to be filled, and is crucial for understanding further the effects of composition on planet properties. Additionally, the topic of the impact of galactic chemical evolution \citep[GCE,][]{Burbidge1957,Lugaro2018} on planet composition has recently joined the conversation in exoplanet science \citep{Frank2014,ONeill2020_GCE}, but stellar compositional scatter is much larger than that predicted by GCE alone. All-in-all, this further warrants a population-level analysis of stellar chemical compositions and, by extension, exoplanetary compositions. Here, we apply the devolatilization trend from \cite{Wang2019} to stellar abundance data from the Hypatia and GALAH catalogues \citep{Hypatia,Buder2018} in an effort to simulate the range of bulk terrestrial exoplanet compositions. The goal is to identify in more detail what compositions needed to be studied, thus providing guidance for an in-depth characterization of rocky exoplanets through future observations, experiments, as well as numerical modellings. 

\section{Data and Methodology} \label{sec:Metho}
We explore terrestrial-type exoplanet compositions by considering rocky planets as devolatilized stars (i.e., stars that lost most of their volatile elements). We use abundances from the Hypatia and GALAH catalogues (Sect.\ \ref{ssec:Metho_stars}). We apply the devolatilization trend from \cite[][Sect.\ \ref{ssec:Metho_star2pl}]{Wang2019} to these abundances to simulate rocky exoplanet compositions. From these compositions, core sizes and mantle compositions are estimated based on a simple model of core-mantle differentiation (Sec.\ \ref{ssec:Metho_coremantle}).

\subsection{Stellar abundances} \label{ssec:Metho_stars}
Stellar abundances from spectroscopy have been summarized in lists such as the Hypatia \citep{Hypatia} and GALAH catalogues \citep{Buder2018}. Constraints from spectroscopy reflect the real stellar composition within a few per cent \citep{Dotter2017}. These observations show that stellar chemical abundances vary significantly from star to star in the solar neighbourhood \citep{Bensby2005,valenti2005,Asplund2009,Lodders2009,Hypatia}, and this chemical diversity should be reflected in the exoplanet population \citep[e.g.,][]{Bond2010,Carter2012,Moriarty2014}.
We firstly obtained stellar abundances from the Hypatia catalogue, an online (routinely updated, retrieved on 2022 July 11th) database compiling measurements from a variety of literature sources \citep{Hypatia}. This database contains elemental abundances of $>9000$ stars within 500 pc from the Sun, which are all of the F, G, K, or M spectral classes. The quality and completeness of abundance data vary per star, but Fe abundances are available for all stars in the database. For converting the available data to molar abundances, and a description of error propagation throughout our work, we refer to the Supplementary Material. For some stars, data from various sources are listed, which can deviate from each other by more than the reported measurement errors. In these cases, we take the median value.

Since data for stars at greater distances are sparse in this catalogue, we only use abundances of those stars within 200 pc (that we here define as the solar neighbourhood). Additionally, we filter for the availability of Mg and Si abundances, as these are the most abundant elements in terrestrial planets, along with Fe. Data on these elements are therefore required for assessing potential terrestrial exoplanet compositions. We also ensure that we only consider stars in the main sequence to prevent systematic errors by filtering for gravitational acceleration of those stars, $\log g > 3.5$. Finally, we exclude data for M-dwarfs, which are yet to be measured accurately due to their faintness and the presence of strong molecular lines in the optical spectra \citep[e.g.,][]{Ishikawa2020}. After these filters, we proceed with abundances for 4236 stars. As previously mentioned, most Sun-like stars are expected to host planets \citep{mulders2018,Bryson2020}, and it is still debatable if the occurrence of rocky planets depends on stellar compositions \citep{Melendez2009,Buchhave2012,Liu2020};
so we do not filter this data for confirmed exoplanet hosts.

\begin{figure*}[h]
   \resizebox{\hsize}{!}{\includegraphics{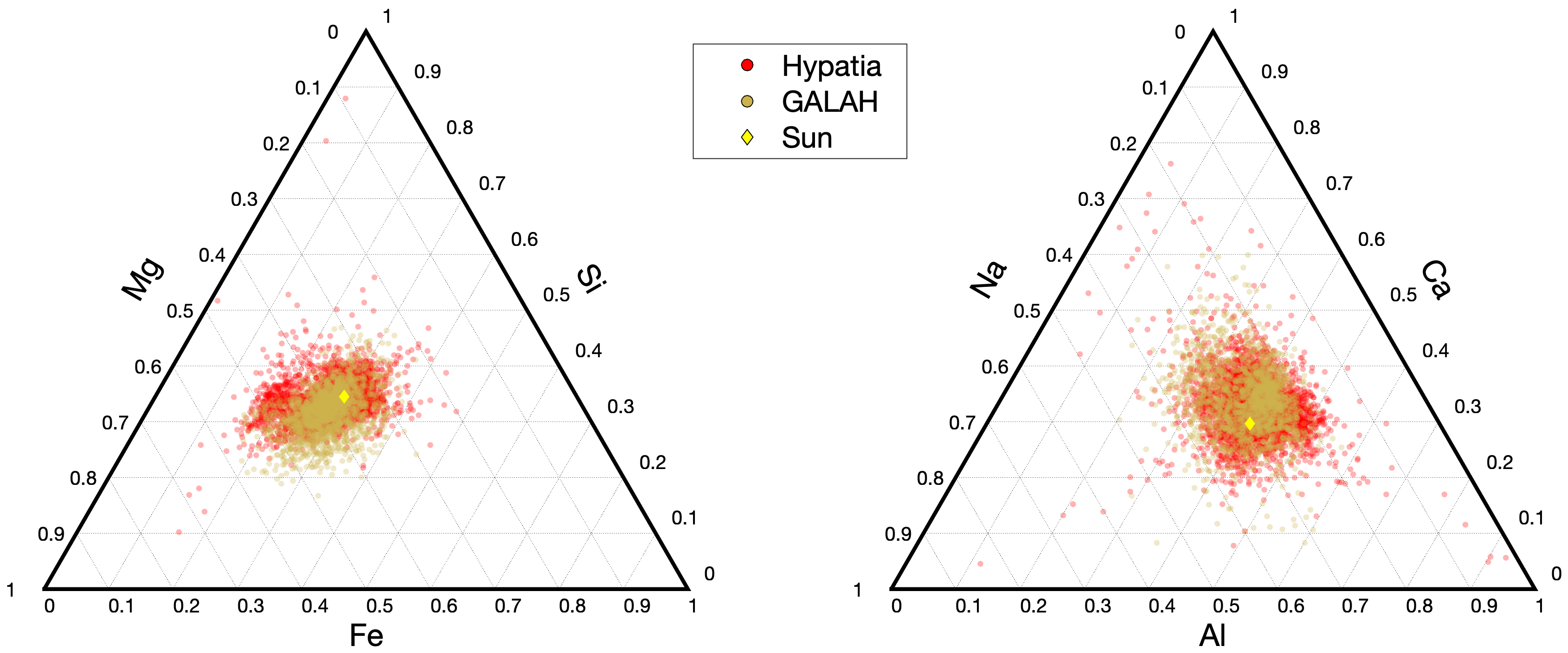}}
   \caption{Stellar abundances documented in the Hypatia (red, 4236 stars) and GALAH (gold, 1971 stars) catalogues used in this study. Solar composition from \cite{Lodders2009} is plotted for comparison.}
   \label{fig:catalogue_comp}
\end{figure*}

Some concerns have been raised regarding systematic errors in the Hypatia catalogue since its data originate from a variety of literature sources with different methodologies, as discussed in \cite{Hinkel2016}. 
To test the robustness of using this catalogue, we consider another inventory of stellar abundances that is intrinsically based on measurements with the same instruments, methodology, and data processing methods. The GALAH catalogue \citep{Buder2018} fits these criteria and is regarded as a reliable data source of stellar abundances. It is also an independent dataset, as it shares only five stars with the Hypatia catalogue \citep{Clark2021}. Target selection for the GALAH survey was based on the 2MASS catalogue \citep{DeSilva2015}, which had no available abundance measurements for Hypatia to include. We compare the data of the two catalogues after applying the same filters described above. Since the GALAH database does not focus on the Solar neighbourhood - the majority of its stars lie further than 200 pc - our filters (including the distance filter) leave just 1971 stars for this catalogue. A comparison of the two catalogues shows that the Hypatia data display a somewhat wider range of compositions, but this may be in part due to the larger number of stars considered (Fig.\ \ref{fig:catalogue_comp}). The spread caused by systematic errors in the Hypatia catalogue seems to be within the bounds of our population spread. Since we find no systematic differences in the datasets of both catalogues (within 200 pc), we merge both datasets and continue with the compositions of 6207 stars.

\subsection{From stellar to planetary abundances} \label{ssec:Metho_star2pl}
We begin with the assumption that stellar abundance data forecast bulk terrestrial exoplanet compositions by presupposing that the observed stellar abundances are a good representation of the planet-forming disc composition. This premise has its basis in the fact that $>$99\% of the mass of any individual star-planet system is that of the star. Further, we assume that terrestrial planet composition is equivalent to the host star composition after applying depletion factors based on the devolatilization trend. This assertion has merit as CI chondritic elemental abundances are a good match for devolatilized solar abundances \citep[for both refractories and moderately volatiles;][]{Anders1989} and for bulk-Earth compositions \citep[for refractories only][]{Wang2019}. We model bulk terrestrial exoplanet compositions based on the Sun-to-Earth devolatilization factors (Tab. \ref{tab:depletion_factors}) from \cite{Wang2019}. We consider key rock-forming elements based on their volatility as represented by the 50\% condensation temperature $T_c$ from \cite{Lodders2003}. We choose to use condensation temperatures from \cite{Lodders2003} as these are the most widely used in the literature, and our analysis has shown that other estimates \citep[e.g.,][]{Wood2019} do not deviate significantly for the elements considered herein. We apply these correction factors (Tab. \ref{tab:depletion_factors}) directly to the stellar abundances and normalize them to yield bulk-planet molar fractions. This application also implicitly confines the planets we consider to the habitable zones around these Sun-like stars. We discuss the applicability and limitations of this devolatilization process in more detail in Sect. \ref{ssec:Disc_devol}.

\begin{table}[h]
\centering
\caption{Depletion factors of all elements we consider here, based on the devolatilization trend from \cite{Wang2019}, alongside condensation temperatures from \cite{Lodders2003}, on which the depletion factors are based.}\label{tab:depletion_factors}
\begin{tabular}{| c | c | c |} 
\hline 
Element & T$_c$ (K) & Depletion factor \\
\hline
O & 875$^{a}$ & 0.810 $\pm$ 0.011 \\
Na & 958 & 0.738 $\pm$ 0.013 \\
Mg & 1336 & 0.148 $\pm$ 0.047 \\
Al & 1653 & 0 \\
Si & 1310 & 0.205 $\pm$ 0.043 \\
S & 664 & 0.929 $\pm$ 0.007 \\
Ca & 1517 & 0 \\
Fe & 1334 & 0.155 $\pm$ 0.047 \\
Ni & 1353 & 0.109 $\pm$ 0.051 \\
\hline
\multicolumn{3}{p{6cm}}{\footnotesize$^a$ The condensation temperature of oxygen was modified by \cite{Wang2019} by considering the nature of oxygen being both a volatile and refractory element.}\\
\end{tabular}
\end{table}

In this work, the elements under consideration are O, Fe, Mg, Si, Ca, Al, Na, S, and Ni; these elements are the most abundant and/or the most critical in terms of their roles in interior properties. The most abundant elements in Solar-System terrestrial planets are Fe, Mg, and Si \citep{Wanke1994,McDonough1995}, and these are also found to be the most abundant refractory elements in all stars in our sample. Additionally, Fe directly affects core size, whereas mantle viscosity depends on the relative abundances of Mg and Si (see below). We include oxygen because of its role in determining core size and forming oxides and silicates. Of the elements not relatively depleted in planets due to their high condensation temperatures, Ca and Al are by far the most abundant, and we therefore consider them. Both elements are likewise important because they influence interior properties, as Ca affects lower mantle mineralogy by stabilising Ca-perovskite, while Al can affect melting behaviour and water storage capacity in minerals. Sodium (Na), while less abundant, significantly affects melting behaviour and crustal buoyancy \citep{Unterborn2017}. Finally, Ni and S affect core size and density, and are therefore also considered.

\subsection{Core-mantle differentiation} \label{ssec:Metho_coremantle}
We separate the modelled bulk planetary compositions into silicate-mantle and metallic-core chemical reservoirs. The size of the core reservoir, and the distribution of Fe between the core and mantle, depends on the amount of oxygen available for oxidation of cations up to iron. However, oxygen availability, expressed as oxygen fugacity (\textit{fO$_2$}), is unfortunately not predictable based on stellar abundances. Oxygen has a dual nature during planet formation as both a volatile (in compounds such as H$_2$O and CO) and as a refractory (in silicates, which are all oxides) element, and its 50\% condensation temperature therefore depends non-linearly on stellar composition \citep[e.g., see][for the effects of Mg and Si]{Unterborn2017_MgSi}. Further, accretion of volatile species increases the bulk planet oxygen fugacity and therefore reduces core size, but accretion of these compounds is not well understood even for the Solar System planets \citep[see e.g.][]{Rubie2015,ONeill2020}. Therefore, we make some assumptions to estimate the amount of oxygen available during core formation. 

Recent studies of polluted white dwarfs show remarkably uniform estimated oxygen fugacities in terrestrial exoplanets \citep{Doyle2019}. This observation leads to the expectation that planets with a similar formation history as Earth would have a similar oxygen fugacity. We therefore model core-mantle differentiation by assuming a fixed Earth-like bulk molar Fe/FeO-ratio, and therefore fixed \textit{fO$_2$} (see Sect.\ \ref{ssec:Disc_coremantle} for a continued discussion). Accordingly, the core size purely depends on the bulk planet Fe abundance, similar to the approach used in \cite{ONeill2020_GCE}. Below, we will mostly consider cases with an Earth-like molar Fe/FeO of 6.31 \citep{McDonough2003,Wang2018_Earthcomp}, but we also consider some cases with variable \textit{fO$_2$} (to be detailed in Sect.\ \ref{ssec:Res_endmembers}). 

In contrast to iron, all Ni and S is assumed to be partitioned into the core due to their high reducing potential and siderophile behaviour \citep{McDonough2003}. We further assume that the core contains 5 wt\% Si \citep{Shahar2009,Javoy2010,Ziegler2010,Hirose2013} and 2.5 wt\% O \citep{McDonough2003,Javoy2010,Hirose2013} as additional light elements, independent of core size and core formation conditions. The light-element budget of the core depends on planet formation scenario and planet size \citep{Wade2005}, amongst other factors, but we only consider planets with one Earth radius, and the resulting differences would not be significant compared to observational errors.

\section{Results} \label{sec:Results}
\begin{figure*}[h]
   \resizebox{\hsize}{!}{\includegraphics{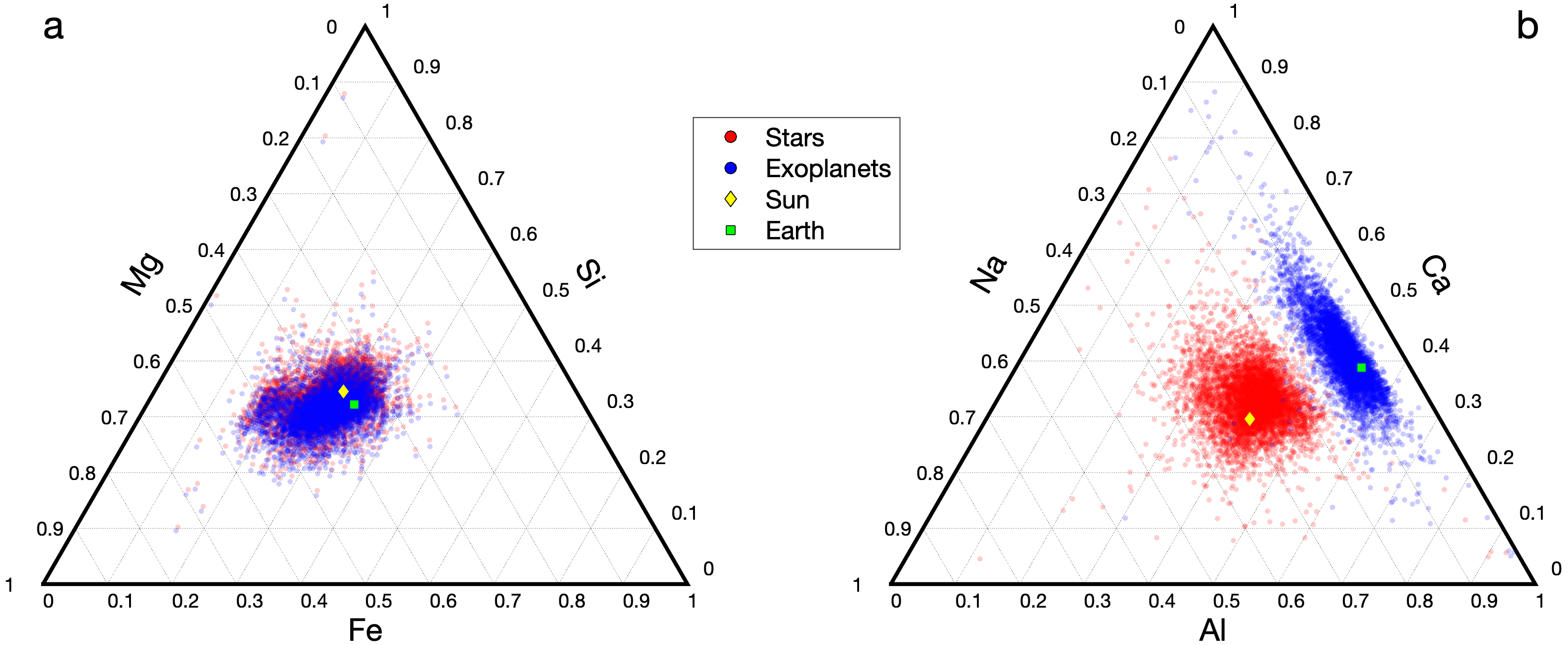}}
   \caption{Bulk compositions of stars in the Solar neighbourhood \citep[red;][]{Hypatia,Buder2018}, planet compositions calculated in this work (blue), Solar composition \citep[yellow diamond;][]{Lodders2009}, and Earth composition \citep[green square;][]{McDonough2003} molar compositions, in the Fe-Mg-Si (left) and Ca-Al-Na (right) systems.}
   \label{fig:Bulk_triangles}
\end{figure*}

We estimate bulk rocky exoplanet compositions in the Solar neighbourhood by applying depletion factors based on the devolatilization trend from \cite{Wang2019} to stellar abundance data in the Solar neighbourhood \citep{Hypatia,Buder2018}. We find that the differences between stellar and planetary compositions are small for the primary rock-forming cations (i.e., major elements on Earth) such as Fe, Mg, and Si (Fig.\ \ref{fig:Bulk_triangles}a), since these elements have similar condensation temperatures in the planet-forming disc. Planets are slightly depleted in Si and Fe relative to more refractory elements (e.g., Ca, Al). In detail, they are more depleted in Si than in Fe, but not by very much. The depletion of terrestrial planets is more pronounced for moderately volatile elements, such as Na (Fig.\ \ref{fig:Bulk_triangles}b), the effect of which is shown and discussed later (Sec.\ \ref{ssec:Res_mineralogy},\ref{ssec:Disc_devol}). In contrast, the highly refractory elements Ca and Al are not depleted in terrestrial planets compared to their host stars.

\subsection{Core sizes} \label{ssec:Res_Core}
\begin{figure}
  \begin{center}
    \includegraphics[width=\textwidth]{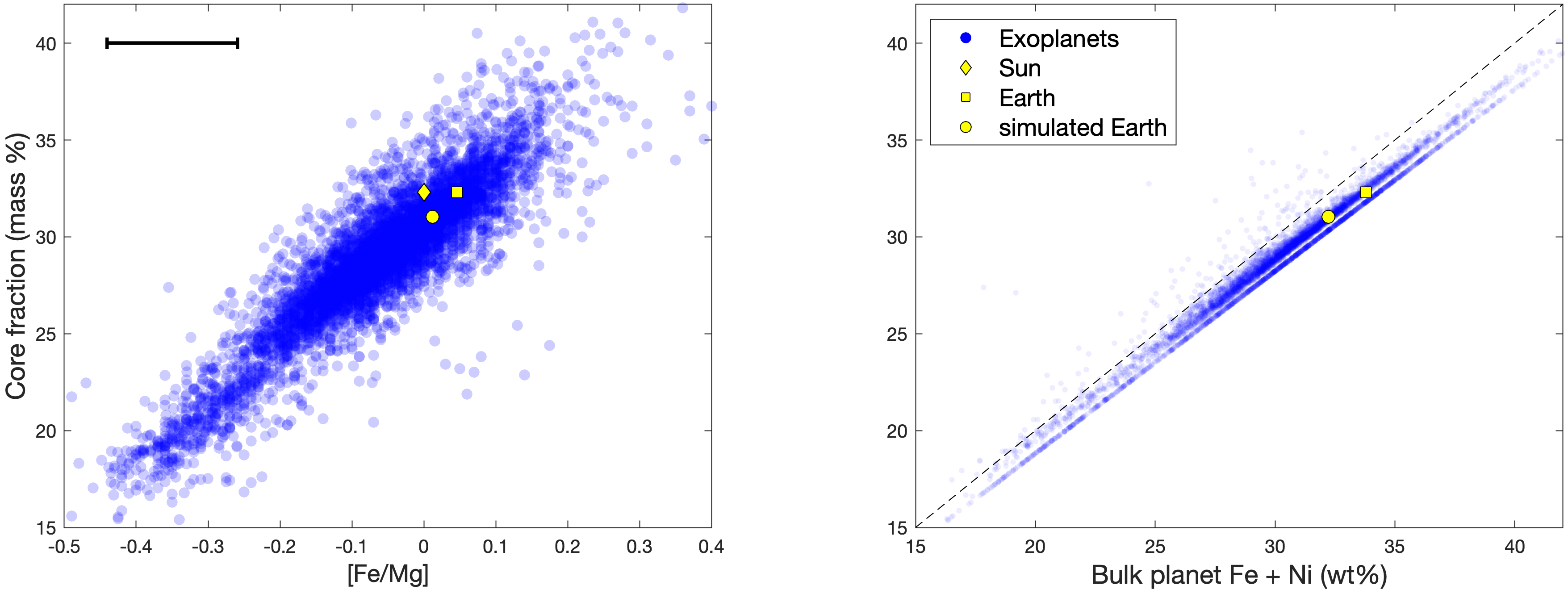}
  \end{center}
   \caption{Core sizes of terrestrial-type exoplanets (mass fraction) as a function of stellar [Fe/Ni] (in dex, left),  and of bulk planet Fe+Ni abundance (wt\%, right). An indication of the typical measurement error on [Fe/Mg] is shown. Core sizes are based on iron distribution between core and mantle according to Earth-like Fe/FeO, and cores containing all Ni and S present in the planet, plus 2.5 wt\% O and 5 wt\% Si. Earth has a core size of 32.3 wt\% for a bulk Fe+Ni content of 33.8 wt\% \citep{McDonough2003}. The dashed line represents the trend for pure Fe-Ni cores where all Fe and Ni is present in the core. The ``simulated Earth'' value is retrieved by applying our methodology to the protosolar composition from \citep{Wang2019}. The Solar value shown is from \cite{Lodders2009}. The GALAH catalogue does not report sulphur abundances, therefore a part of our population has noticably smaller cores due to the absence of sulphur.}
   \label{fig:Core_masses}
\end{figure}

To understand long-term planetary evolution and interior-exterior coupling, it is important to constrain mantle compositions based on bulk planet composition. Core-mantle differentiation is highly efficient for most elements, which are effectively partitioned into either the core or mantle. In contrast, Fe can enter both the mantle (as iron oxides and in silicates) and core (as Fe$^0$), controlled by the availability of oxygen (i.e., the oxygen fugacity). We assume that iron partitions between the core and mantle in a similar way as on Earth (i.e., Fe/FeO is equal to that of Earth, see Sect.\ \ref{ssec:Metho_coremantle}). This assumption leads to a nearly linear trend of core size with bulk planet Fe abundance (Fig.\ \ref{fig:Core_masses}). Further, bulk planet Fe abundance is a function of stellar [Fe/Mg], and therefore a simple linear trend exists between stellar [Fe/Mg] and core mass. Deviation from linearity in either trend stems from light elements in the core, as we assume that all S is partitioned into the core. Based on our model, Earth's core size is close to the population average. While our simulated Earth composition (based on the Proto Solar composition from \cite{Wang2019}) can deviate from the composition presented by \cite{McDonough2003} (Fig.\ \ref{fig:Core_masses}), there are multiple estimates in the literature that deviate from theirs \citep[e.g.,][]{Wang2018_Earthcomp}. Further, the measurement errors on stellar abundances are typically large enough to allow for these deviations. 

Variation in stellar S content leads to a diversity in core light element fraction among rocky exoplanets, under our assumption that all S is partitioned into the core. All exoplanets in our population have core light element fractions of at least 7.5 wt\%, as we assume that rocky planet cores contain 5 wt\% Si and 2.5 wt\% O, similar to Earth's core. Further addition of S increases the light element fraction to between 8 and 12 wt\%. (Fig.\ \ref{fig:Suppl_Core_comp}a). There seems to be a minor trend where core S fraction is slightly higher for smaller cores (Fig.\ \ref{fig:Suppl_Core_comp}b), so small cores are also slightly less dense than large cores. Further, core Ni content can vary from 5 to 9 mol\% (cf.\ 6\% in Earth), having limited effect on core sizes (Fig.\ \ref{fig:Suppl_Core_comp}d).

\subsection{Mantle compositions} \label{ssec:Res_Mantle}
\begin{figure*}
    \resizebox{\hsize}{!}{\includegraphics{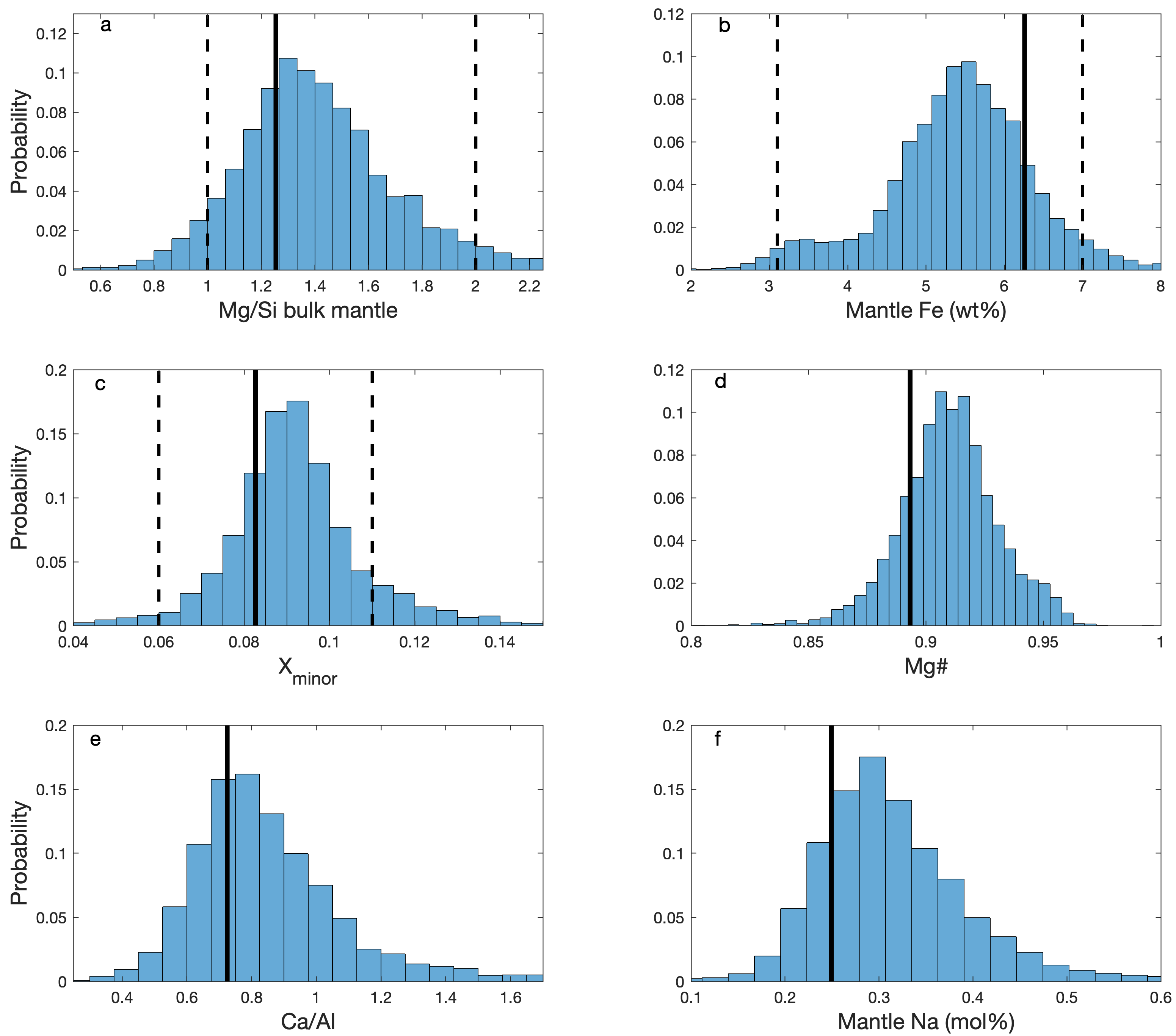}}
      \caption{Histogram of terrestrial-type exoplanet mantle compositions, in terms of molar Mg/Si-ratios (a), iron abundance in wt\% (b), molar fraction of minor elements (i.e., Ca, Al, and Na; c), molar Mg number (MgO/(MgO+FeO); d), molar Ca/Al-ratios (e), and mantle molar Na abundance (f). For comparison, the Earth composition is plotted as a solid black line \citep[Mg/Si=1.25, 6.26 wt\% Fe, 8.21\% minor elements, Mg\#=0.89, Ca/Al=0.73, 0.25 mol\% Na,][]{McDonough2003}. The synthetic representative compositions (Sect.\ \ref{ssec:Res_endmembers}) are shown in panels a, b, and c, as dashed lines.}
      \label{fig:Mantle_histograms}
\end{figure*}

We estimate rocky exoplanet mantle compositions by subtracting the simulated core size and composition from the bulk compositions. The most abundant cations in the Earth's mantle are Fe, Mg, and Si, dubbed major elements. The other elements (aside from O; Ca, Al, Na, S, and Ni) are referred to as minor elements below. The abundances of Mg and Si are often conveyed as the Mg/Si ratio, as this is an indicator for relative abundances of common mantle minerals, such as olivine ((Mg,Fe)$_2$SiO$_4$) and pyroxene ((Mg,Fe)SiO$_3$) in the upper mantle, or bridgmanite ((Mg,Fe)SiO$_3$) and ferropericlase ((Mg,Fe)O) in the lower mantle. Further, it indicates the appearances of minerals which are rare in Earth, such as ferropericlase in the upper mantle at high Mg/Si, or quartz at low Mg/Si (see Sects.\ \ref{ssec:Res_mineralogy}, \ref{ssec:Disc_comp}). Most planets we consider here have mantle Mg/Si-ratios between 0.8 and 2.2 (see Fig.\ \ref{fig:Mantle_histograms}a). The bulk silicate Earth with a Mg/Si-ratio of 1.2 may therefore be viewed as below average. Notable is that planet mantle Mg/Si is typically 15-20\% higher than stellar Mg/Si, with about a 7\% increase due to the difference in depletion factors (see Table \ref{tab:depletion_factors}), and a further depletion due to the presence of Si in the core.

The planets in our population also have mantle iron contents of 3 - 7 wt\% (Fig.\ \ref{fig:Mantle_histograms}b). The Earth is above average in terms of mantle iron content, at 6.26 wt\%. This is also reflected in the mantle Mg\# (molar Mg/(Mg+Fe)), where the Earth (0.89) is accordingly below average (0.87-0.95; Fig.\ \ref{fig:Mantle_histograms}d).

The abundance of minor elements is an indicator of how exotic an exoplanet composition is. The minor elements, Ca, Al, and Na, are significantly less abundant than Mg, Fe and Si in the Solar System planets. Here, we define minor element fraction as $X_{minor} = (Ca+Al+Na)/(Ca+Al+Na+Fe+Mg+Si)$. In our terrestrial-type exoplanet population, $X_{minor}$ ranges from 6 to 13 molar per cent (see Fig.\ \ref{fig:Mantle_histograms}c). Therefore, we find that exotic compositions with extreme abundances of minor elements are rare, consistent with inferences from GCE \citep{Alibes2001}. Relative abundances of minor elements vary, as illustrated by the molar Ca/Al ratios, which show a significant skew (i.e., tail at high Ca/Al; see Fig.\ \ref{fig:Mantle_histograms}e). The Na mantle abundance shows a similar shape to the Ca/Al distribution and is consistently low for all exoplanet compositions (Fig.\ \ref{fig:Mantle_histograms}f). From our analysis, we can conclude that Earth is close to the median value in terms of total minor element abundance, Ca/Al-ratio, and Na abundance.  

\begin{table}[h]
\centering
\caption{Statistics of terrestrial-type exoplanet compositions in the Solar neighbourhood. Mantle (top half) and core (bottom half) compositions are in mol\% (or molar ratios) unless specified otherwise. Values for Earth from \cite{McDonough2003} are given for comparison. Mg$\#$ is defined as molar mantle Mg/(Mg+Fe).} \label{tab:comp_statistics}
\begin{tabular}{| c | c | c | c |} 
\hline 
Parameter & Median & 2$\sigma$ range & Earth \\
\hline
Mg/Si & 1.39 & (0.78, 2.18) & 1.25 \\
Mantle Fe (wt\%) & 5.48 & (3.17, 7.57) & 6.26 \\ 
Mg\# & 0.91 & (0.86, 0.95) & 0.89 \\
$X_{minor}$ & 9.24 & (4.99, 13.66) & 8.21 \\ 
Ca/Al & 0.88 & (0.45, 1.62) & 0.73 \\
Na & 0.321  & (0.183, 0.549) & 0.25 \\ 
\hline
Core mass (wt\%) & 28.48 & (18.94, 35.57) & 32.3 \\ 
Core S & 4.25 & (1.99, 10.96) & 1.9 \\ 
\hline
\end{tabular}
\end{table}

\subsection{Representative compositions} \label{ssec:Res_endmembers}
Next, we select 21 representative compositions for our population of modelled rocky exoplanets. These compositions are chosen such that they span the compositional range inferred by our approach, and also capture key trends in the dataset. We study these representative compositions further by simulating mantle mineralogy of each of these model planets. These representative compositions may be used as a convenient reference dataset for further studying the effects of rocky exoplanet composition.

The terrestrial-type exoplanet mantle compositions we retrieve can be described to first order by normal distributions (Fig. \ref{fig:Mantle_histograms}). Table \ref{tab:comp_statistics} provides the average value and 2-sigma range of key compositional properties. However, the dataset also displays systematic compositional trends that are not captured by simple normal distributions (Fig.\ \ref{fig:Comp_scatter}), likely due to chemical trends in the galaxy \citep[e.g.,][]{Bensby2011}. For example, planets with high core mass tend to have lower Mg/Si (see Fig.\ \ref{fig:Comp_scatter}a), as stellar Mg/Si tends to decrease with increasing iron abundance due to GCE (see Sect.\ \ref{ssec:Disc_Stars}). 
For a full overview of all compositional trends, please refer to Fig.\ \ref{fig:Suppl_scatter}. 

Because of the presence of compositional trends, we define a number of representative compositions that capture these trends to study in more detail, rather than studying planet compositions at the 2$\sigma$ limits.
Our 21 representative compositions represent the full compositional range (within 95\% confidence levels) of the distribution of our modelled exoplanets 
(Tab.\ \ref{tab:comp_statistics}). We arrive at these 21 representative planet compositions by combining three approaches. First, we pick 8 actual samples from our (devolatilized) dataset. These \textit{sample-based} compositions are chosen such as to span the compositional range for all parameters. Next, we define 9 \textit{synthetic} planet compositions based on the 2 sigma limits of our exoplanet distribution in terms of mantle Mg/Si, core size, and bulk planet Fe+Mg+Si, to study the effects of varying Fe, Mg, and Si individually, as these are the most abundant cations in all exoplanets. Finally, we define 4 additional synthetic planet compositions, dubbed \textit{synthetic-fO$_2$}, with different bulk \textit{fO$_2$} and therefore different Fe/FeO, to test our assumption of Earth-like Fe/FeO. The mantle compositions and core sizes of these 21 representative compositions can be found in Table \ref{tab:endmember_comps}.

\subsubsection{Sample-based compositions}\label{sssec:Res_comp_sample}
To begin with, eight \textit{sample-based} compositions are selected from our simulated planet population, earmarked by applying the selection criteria below until only a few planets remain, at which point the best sample is selected by hand. We summarize these selection criteria both in text, and in Table \ref{tab:comp_method_choice}. These \textit{sample-based} compositions are divided into high-Fe/Mg and low-Fe/Mg planets, and we define four compositions for each of these subsets. 
Planets with high Fe/Mg exhibit the highest core masses and lowest Mg/Si-ratios. Thus, we select one composition with high core mass fraction and relatively high Mg/Si (\textit{Sample high M$_c$}), and one composition with low Mg/Si and relatively low core mass fraction (\textit{Sample low Mg/Si}). High core mass fractions are typically associated with low Na abundances (Fig.\ \ref{fig:Comp_scatter}c), as well as average Ca/Al and minor element fractions (Fig.\ \ref{fig:Suppl_scatter}), so \textit{Sample high M$_c$} represents these trends. 
Low Mg/Si is typically associated with low minor element fractions, and average Na abundances (Fig.\ \ref{fig:Suppl_scatter}): \textit{sample low Mg/Si} is chosen to reflect these characteristics. We further take a composition with high Ca/Al-ratios (\textit{Sample high Ca/Al}) and another one with high minor element abundances (\textit{Sample high minor}), as these characteristics are typically related to high-Fe/Mg compositions (Figs.\ \ref{fig:Comp_scatter}b, \ref{fig:Suppl_scatter}). The former is associated with lower-than-average minor element fractions (Fig.\ \ref{fig:Comp_scatter}d). The latter is associated with low Ca/Al (Fig.\ \ref{fig:Comp_scatter}d) and average Mg/Si (Fig.\ \ref{fig:Comp_scatter}b).

Planets with low Fe/Mg exhibit the lowest core masses and highest Mg/Si-ratios. To reflect these systematics, two planets are chosen based on low core mass (\textit{Sample low M$_c$}) and high Mg/Si (\textit{Sample high Mg/Si}), respectively. The former is associated to the lowest Na abundances (Fig.\ \ref{fig:Comp_scatter}c), and average Ca/Al and minor elements (Fig.\ \ref{fig:Suppl_scatter}). The latter is associated with below-average Ca/Al-ratios (Fig.\ \ref{fig:Comp_scatter}b), but average minor element fractions and Na abundances (Fig.\ \ref{fig:Suppl_scatter}). Further, we include a planet with high Na-abundances among the low-Fe/Mg cases (\textit{Sample high Na}), which corresponds to a high core mass fraction for the low-Fe/Mg planets (Fig.\ \ref{fig:Comp_scatter}b), as well as fairly high Ca/Al. Finally, we choose a composition with very low minor element fractions (\textit{Sample low minor}), which correspond to low Mg/Si-ratios (Fig.\ \ref{fig:Suppl_scatter}).

\definecolor{very_high}{RGB}{127,178,115}
\definecolor{high}{RGB}{170,220,120}
\definecolor{low}{RGB}{150,200,210}
\definecolor{very_low}{RGB}{100,160,180}

\begin{table*}[t]
\centering
\caption{Compositional trends on which the eight sample-based representative compositions have been chosen. Trends are defined based on Figures \ref{fig:Comp_scatter} and \ref{fig:Suppl_scatter}. }\label{tab:comp_method_choice}
\begin{tabular}{| c | c | c | c | c | c | c  |} 
\hline 
Defining characteristic & Fe/Mg & Typical M$_c$ & Typical Mg/Si & Typical minor fraction & Typical Ca/Al & Typical Na \\
\hline
Sample high M$_c$ & \cellcolor{high} High & \cellcolor{very_high} Very high & Average & Average & Average & \cellcolor{low} Low \\
Sample low Mg/Si & \cellcolor{high} High & Average & \cellcolor{very_low} Very low & \cellcolor{low} Low & Average & Average \\
Sample high minor & \cellcolor{high} High & Average & Average & \cellcolor{very_high} Very high & \cellcolor{low} Low & \cellcolor{very_high} Very high \\
Sample high Ca/Al & \cellcolor{high} High & \cellcolor{high} High & Average & \cellcolor{low} Low & \cellcolor{very_high} Very high & Average \\
\hline
Sample low M$_c$ & \cellcolor{low} Low & \cellcolor{very_low} Very low & \cellcolor{high} High & Average & Average & \cellcolor{very_low} Very low \\
Sample high Mg/Si & \cellcolor{low} Low & \cellcolor{low} Low & \cellcolor{very_high} Very high & Average & \cellcolor{low} Low & Average \\
Sample high Na & \cellcolor{low} Low & \cellcolor{low} Low & \cellcolor{high} High & \cellcolor{low} Low & \cellcolor{high} High & \cellcolor{high} High \\
Sample low minor & \cellcolor{low} Low & \cellcolor{low} Low & \cellcolor{low} Low & \cellcolor{low} Low & Average & \cellcolor{low} Low \\
\hline
\end{tabular}
\end{table*}

\subsubsection{Synthetic compositions}\label{sssec:Res_comp_synth}
Next, eight of the nine \textit{synthetic} compositions are based solely on varying core size, mantle Mg/Si, and mantle minor element fractions. These quantities together span the bulk planet Fe, Mg, and Si fractions and have the most significant effect on interior properties. For other elements, their fractions in the \textit{synthetic} compositional cases are assumed to be equal to Earth, allowing us to study the effects of varying Fe, Mg, and Si alone. We use mantle molar Mg/Si, with upper and lower bounds 1.0 and 2.0 (Fig.\ \ref{fig:Mantle_histograms}). Further, we use core mass fractions of 20 and 35 wt\%. This also defines mantle Fe fraction, as we assume Earth-like core S content and \textit{fO$_2$} is assumed constant, to between 1.3 and 2.8 mol\% (corresponding to mantle FeO weight percentages of about 3.1 to 7.0). Finally, we define the synthetic compositions using mantle minor element fractions of 6 and 11 mol\%, as this also determines the bulk mantle Fe+Mg+Si content. We include a ninth \textit{synthetic} composition based on the median values of core size, mantle Mg/Si, and minor element fraction (See Tab.\ \ref{tab:comp_statistics}). Figure \ref{fig:Suppl_scatter} shows that the 9 synthetic and the 8 sample-based compositions cover the full range of inferred rocky exoplanet compositions.

\begin{figure*}
    \resizebox{\hsize}{!}{\includegraphics{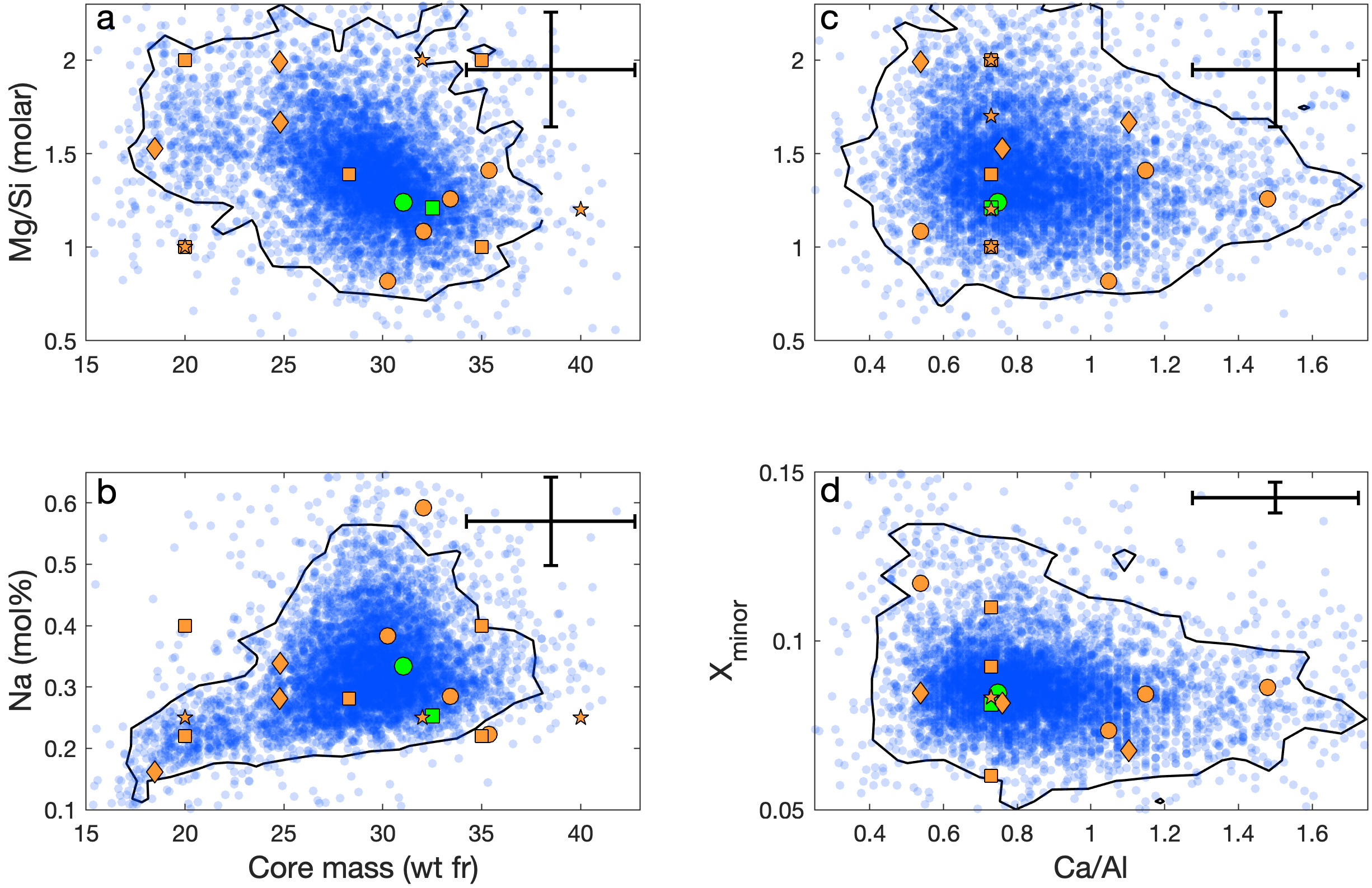}}
      \caption{Scatter plots of bulk terrestrial exoplanet compositions for a few chosen compositional quantities, showing compositional trends. The contours contain 95\% of the modelled compositions. Orange dots and diamonds show the high-Fe/Mg and low-Fe/Mg \textit{sample-based} compositions, respectively, the orange squares show the eight \textit{synthetic} compositions, and the orange stars show the four \textit{synthetic-fO$_2$} compositions. Note that composition \textit{synthetic-fO$_2$} 4 falls outside the range plotted in panels (a) and (c) (see Figure \ref{fig:Suppl_scatter}). The light green square represents Earth composition \citep{McDonough2003}, while the light green circle represents the composition of a simulated planet based on a star with a protosolar composition \citep{Wang2019}. Error-bars are based on combined uncertainties of stellar abundance measurements from the Hypatia and GALAH catalogues \citep{Hypatia,Buder2018} and of the devolatilization trend (see Table \ref{tab:depletion_factors}).}
      \label{fig:Comp_scatter}
\end{figure*}

\subsubsection{fO2-based compositions}\label{sssec:Res_comp_oxfug}
While these seventeen compositions are selected based on the assumption of Earth-like \textit{fO$_2$} during core formation, we also consider four cases with variable \textit{fO$_2$} for comparison. To constrain these four compositions, we recalculate core-mantle differentiation for all of our hypothetical 6207 planets based on observed stellar oxygen-to-refractory-element ratios (as opposed to the assumption of a fixed Fe/FeO as above). Applying the Earth-Sun oxygen devolatilization factor results in planets with oxygen to cation ratios between 0.5 and 2.06, where 1.0 means sufficient oxygen is available to oxidize all cations to oxides, including all metallic iron. However, we expect that coreless exoplanets are rare, as planetary material accreting onto polluted white dwarf stars typically comes from planets that differentiated into a mantle and core \citep[e.g.,][]{Hollands2018,Doyle2019,Bonsor2020}. Further, we consider that planets with elements other than Fe and Ni that are reduced to a metallic state due to exceptionally low oxygen budget are also rare. Therefore, we assume that the oxygen depletion factor scales with stellar oxygen-to-refractory-element ratios in such a way that both coreless and extremely reduced (with other cations than Fe and Ni in the core) fall on the edges of the 95\% confidence level, and so the distribution accurately reproduces the Earth-Sun oxygen depletion factor (Table \ref{tab:depletion_factors}). We choose the four compositions near the edges of the 95\% confidence level of the remaining planetary oxygen budget distribution (Suppl.\ Fig.\ \ref{fig:Suppl_Fe_FeO}), as well as upper and lower bulk planet Mg/Fe values. This approach results in planet Fe/FeO values between 0.16 and 56.4.

\begin{table*}[h]
\centering
\caption{Core sizes (weight per cent) and mantle oxide compositions (molar per cent) of the 21 representative compositions.}\label{tab:endmember_comps}
\begin{tabular}{| c | c | c | c | c | c | c | c |} 
\hline 
Composition & M$_c$ & Na$_2$O & MgO & Al$_2$O$_3$ & SiO$_2$ & CaO & FeO \\
\hline
Earth & 32.5 & 0.30 & 48.24 & 2.23 & 39.85 & 3.25 & 5.96 \\
\hline
Sample high M$_c$ & 35.1 & 0.27 & 50.48 & 1.87 & 35.76 & 4.29 & 6.90 \\
Sample low Mg/Si & 30.1 & 0.42 & 40.11 & 1.59 & 48.92 & 3.33 & 5.46 \\
Sample high minor & 31.4 & 0.75 & 44.66 & 3.48 & 41.17 & 3.74 & 6.12 \\
Sample high Ca/Al & 33.3 & 0.35 & 48.29 & 1.63 & 38.40 & 4.82 & 6.42 \\
Sample low M$_c$ & 18.2 & 0.20 & 55.03 & 2.26 & 36.06 & 3.43 & 2.78 \\
Sample high Mg/Si & 24.4 & 0.33 & 59.86 & 2.60 & 30.08 & 2.80 & 3.90 \\
Sample high Na & 24.6 & 0.40 & 56.77 & 1.44 & 34.06 & 3.17 & 3.92 \\
Sample low minor & 20.3 & 0.26 & 49.49 & 1.30 & 42.79 & 2.78 & 3.38 \\
\hline
Synthetic 1 & 35 & 0.42 & 56.91 & 2.98 & 28.45 & 4.34 & 6.67 \\
Synthetic 2 & 35 & 0.22 & 59.43 & 1.60 & 29.73 & 2.33 & 6.58 \\
Synthetic 3 & 35 & 0.42 & 42.49 & 2.98 & 42.49 & 4.34 & 7.06 \\
Synthetic 4 & 35 & 0.22 & 44.36 & 1.60 & 44.36 & 2.35 & 6.99 \\
Synthetic 5 & 20 & 0.42 & 59.26 & 2.98 & 29.63 & 4.36 & 3.11 \\
Synthetic 6 & 20 & 0.22 & 61.76 & 1.61 & 30.88 & 2.34 & 3.07 \\
Synthetic 7 & 20 & 0.42 & 44.36 & 2.99 & 44.36 & 4.34 & 3.30 \\
Synthetic 8 & 20 & 0.23 & 46.22 & 1.60 & 46.22 & 2.34 & 3.27 \\
Synthetic 9 & 28.3 & 0.34 & 51.0 & 3.37 & 36.66 & 3.69 & 4.99 \\
\hline
Synthetic-fO$_2$ 1 & 40 & 0.31 & 50.60 & 2.23 & 42.17 & 3.27 & 1.25 \\
Synthetic-fO$_2$ 2 & 32 & 0.31 & 61.89 & 2.24 & 30.93 & 3.27 & 0.24 \\
Synthetic-fO$_2$ 3 & 20 & 0.32 & 38.40 & 2.24 & 38.40 & 3.27 & 17.20 \\
Synthetic-fO$_2$ 4 & 7 & 0.32 & 50.24 & 2.24 & 29.56 & 3.27 & 14.21 \\
\hline
\end{tabular}
\end{table*}

\subsection{Mantle mineralogy} \label{ssec:Res_mineralogy}
\begin{figure*}
    \resizebox{\hsize}{!}{\includegraphics{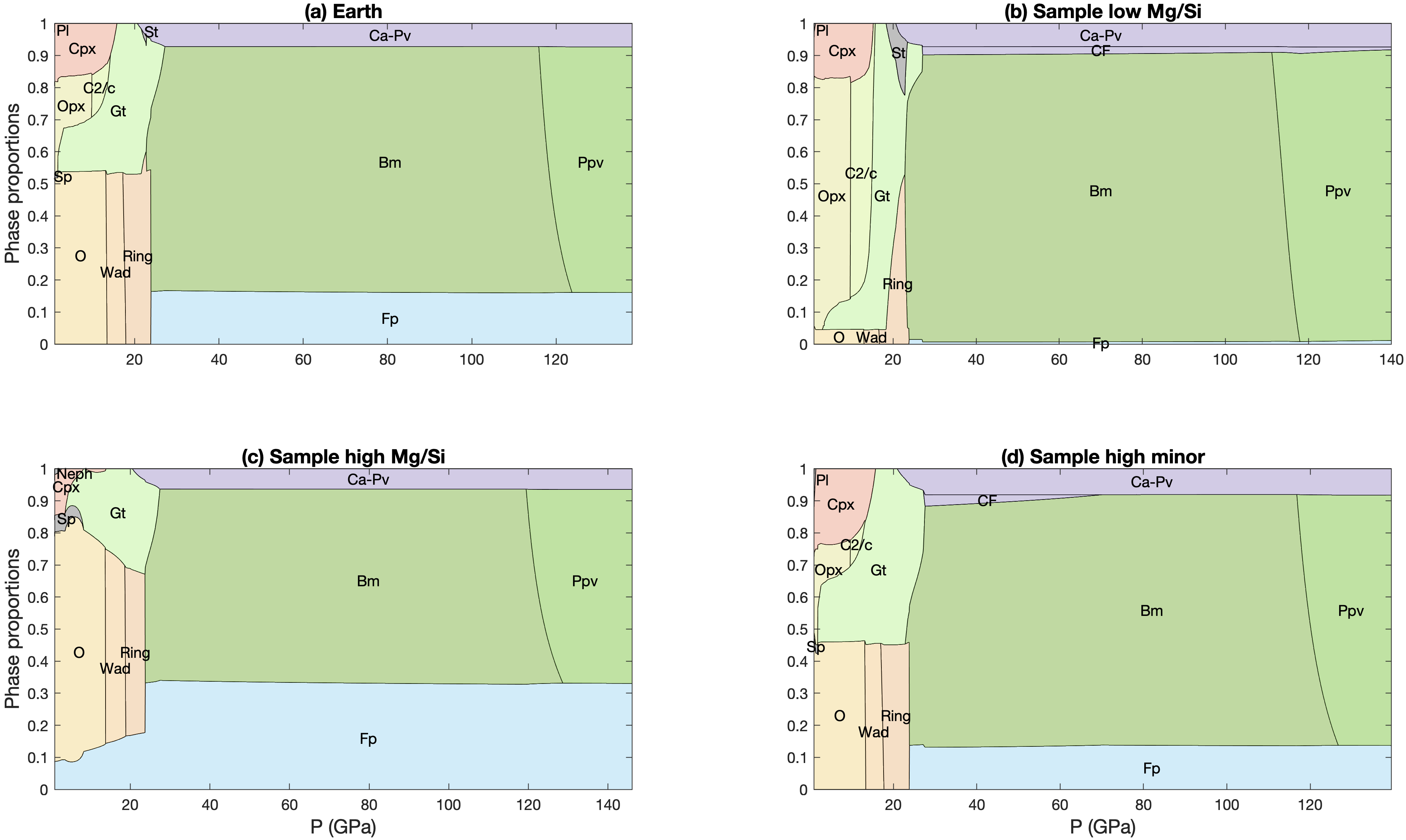}}
      \caption{Mineralogical profiles of Earth \citep[composition from][]{McDonough2003} and three selected representative compositions (see Table \ref{tab:endmember_comps}), showing mineralogical diversity. Pressure is shown from the surface (left side of each profile) to the core-mantle boundary pressure (right side), determined from Eq.\ 16 of \cite{Noack2020}. Mineralogical assemblages are calculated with Perple\_X \citep{Connolly2005}, using the thermodynamic database from \cite{Stixrude2022}. O:olivine, Opx:orthopyroxene, Cpx:clinopyroxene, C2/c:C2/c pyroxene, Neph:Nepehline, Q:Quartz, Pl:Plagioclase, Sp:Spinel, Gt:Garnet, Wad:Wadsleyite, Ring:Ringwoodite, St:Stishovite, Bm:bridgmanite, Fp:ferropericlase, Ca-Pv:Ca-perovskite, CF:Ca-ferrite, Ppv:post-perovskite.}
      \label{fig:Mineralogy}
\end{figure*}

Finally, we model mantle mineralogical profiles for each of these 21 representative planet compositions. To this end, we employ a Gibbs free energy minimisation algorithm, Perple\_X \citep{Connolly2005}, using the thermodynamic database from \cite{Stixrude2022}. This algorithm estimates the stable mantle mineralogy as a function of pressure and temperature. We simulate mantle mineralogy based on the present-day Earth mantle adiabat \citep{Brown1981}, up to the core-mantle boundary pressure of any given planet, which depends on core size and bulk iron abundance (according to Eq.\ 16 from \citealt{Noack2020}). Generally, the mineralogical profiles of the representative cases are similar to those of Earth, with the upper mantle mainly consisting of olivine (and its high-pressure polymorphs wadsleyite and ringwoodite; (Mg,Fe)$_2$SiO$_4$), pyroxene ((Mg,Fe)SiO$_3$), and garnet ((Ca,Mg,Fe)$_3$(Al,Fe)$_2$Si$_3$O$_{12}$), and the lower mantle of bridgmanite and post-perovskite ((Mg,Fe)SiO$_3$), ferropericlase ((Mg,Fe)O), and Ca-perovskite (CaSiO$_3$). However, the relative abundances of these minerals vary significantly among these planets (Fig.\ \ref{fig:Mineralogy}). 

\begin{figure*}[h]
    \resizebox{\hsize}{!}{\includegraphics{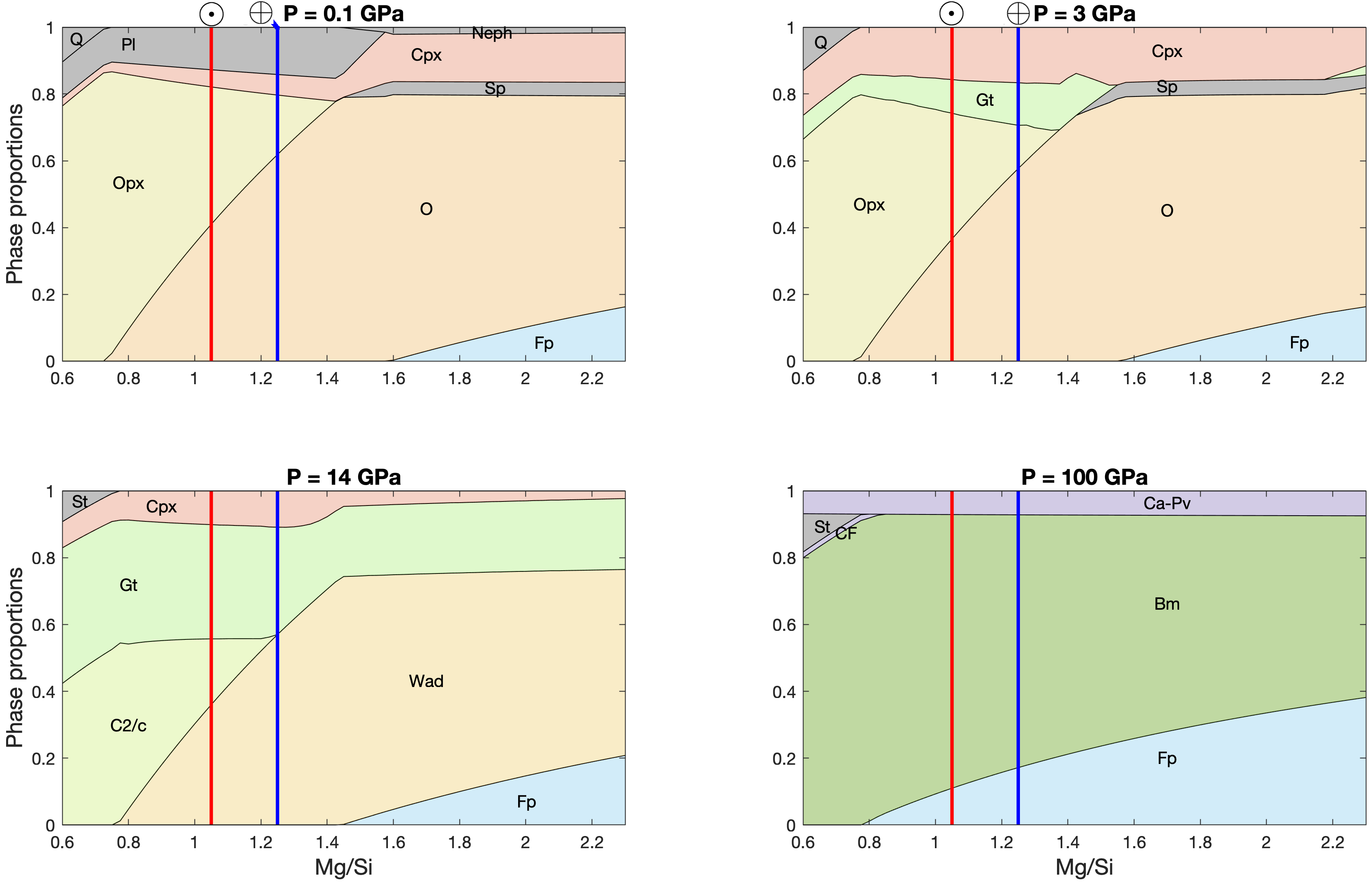}}
      \caption{Mantle mineralogy at a range of pressures for planets with Earth-like composition \citep[composition from][]{McDonough2003} in terms of Al$_2$O$_3$, CaO, and Na$_2$O, and Mg/Si varying from 0.6 to 2.0. The FeO abundance is increased linearly with decreasing Mg/Si from 4 to 7 mol\%. Mineralogical assemblages are calculated with Perple\_X \citep{Connolly2005}, using the thermodynamic database from \cite{Stixrude2022}. Solar \citep[red, $\odot$; ][]{Lodders2009} and Earth \citep[blue, $\oplus$; ][]{McDonough2003} compositions are indicated in all panels. Mineral abbreviations are given in Figure \ref{fig:Mineralogy}.}
      \label{fig:MgSi_mineralogy}
\end{figure*}

In detail, Earth is dominated by olivine and bridgmanite, with pyroxene, garnet, and ferropericlase as the main secondary phases (Fig.\ \ref{fig:Mineralogy}a). In planets with low Mg/Si (Fig.\ \ref{fig:Mineralogy}b), the olivine fraction is significantly lower, and the upper mantle consists primarily of pyroxene and garnet. Further, ferropericlase abundances in the lower mantle of these planets approach zero. In turn, planets with high Mg/Si contain significantly more olivine and ferropericlase in their mantles (Fig.\ \ref{fig:Mineralogy}c). Further, the most Mg-enriched planets contain ferropericlase in the upper mantle, even at very low pressures. Planets with sufficiently high Mg/Si to contain ferropericlase in the upper mantle also completely lack orthopyroxene, while clinopyroxene and garnet are still present to accommodate Na, Al, and Ca. Finally, planets with high minor element abundances have higher abundances of minerals such as Ca-perovskite, Ca-ferrite, and clinopyroxene (Fig.\ \ref{fig:Mineralogy}d). However, even in these planets, the primary phases remain the Fe,Mg-silicates olivine, pyroxene, bridgmanite, and ferropericlase.

\begin{figure*}[h]
    \resizebox{\hsize}{!}{\includegraphics{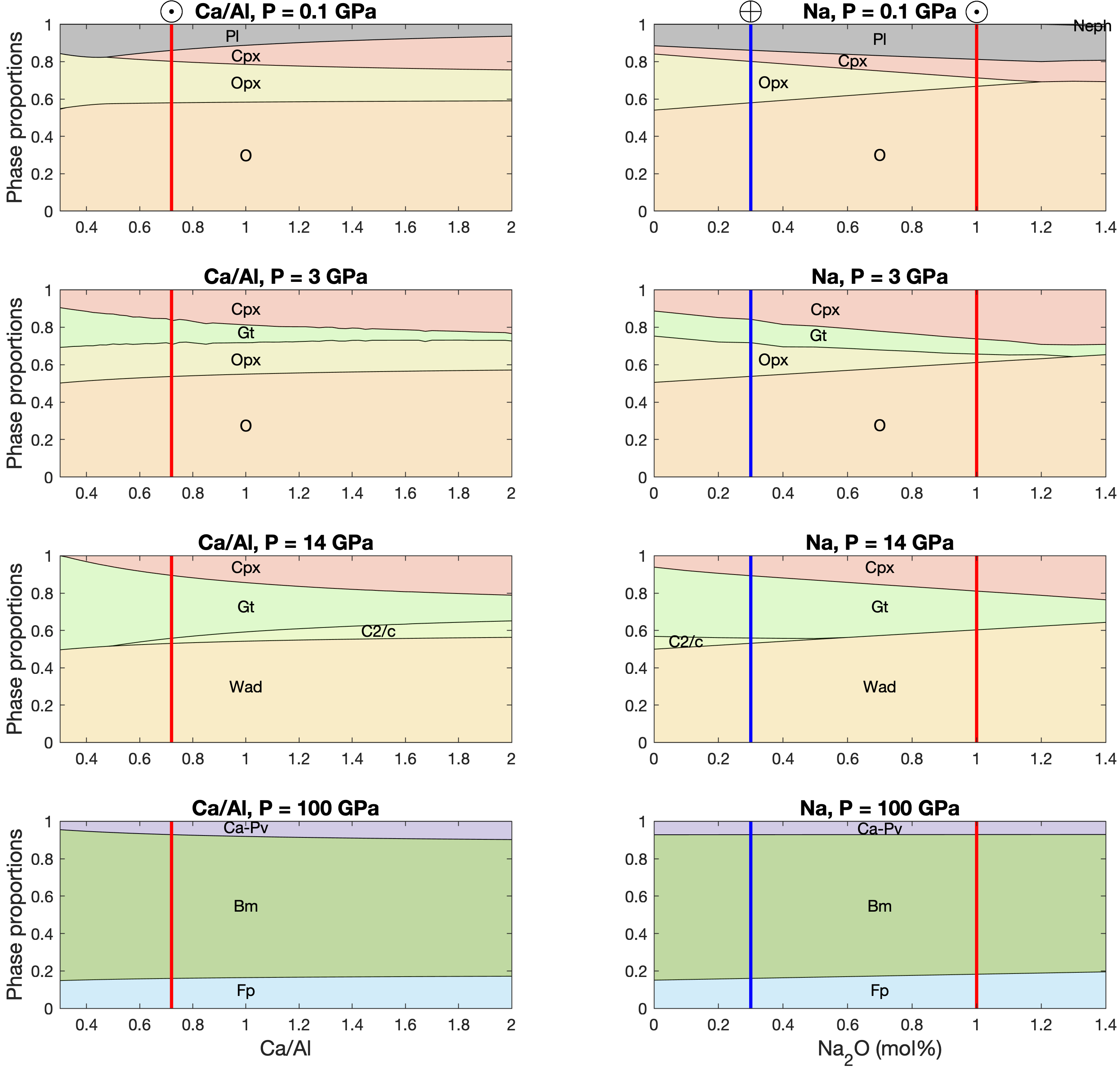}}
      \caption{Mantle mineralogy at a range of pressures for planets with Earth-like composition \citep[composition from][]{McDonough2003} in terms of FeO, MgO, and SiO$_2$, and Ca/Al varying from 0.3 to 1.7 (left), and Na$_2$O abundances ranging from 0 to 1 mol\% (left). Mineralogical assemblages are calculated with Perple\_X \citep{Connolly2005}, using the thermodynamic database from \cite{Stixrude2022}. Solar \citep[red, $\odot$; ][]{Lodders2009} and Earth \citep[blue, $\oplus$; ][]{McDonough2003} compositions are indicated in all panels (Solar and Earth Ca/Al coincide). Mineral abbreviations are given in Figure \ref{fig:Mineralogy}.}
      \label{fig:CaAlNa_mineralogy}
\end{figure*} 

As seen in the mineralogical profiles of our representative compositions, Mg/Si is an important control on the main mantle mineralogy. Assuming otherwise Earth-like composition shows important transitions in upper mantle mineralogy, where for Mg/Si $\geq$ 1.6 ferropericlase becomes present, while for Mg/Si $\leq$ 0.8 the SiO$_2$-minerals quartz and stishovite appear (Fig.\ \ref{fig:MgSi_mineralogy}). Note that quartz-bearing mantles fall just outside of our compositional spread (Fig.\ \ref{fig:Mantle_histograms}a). Further, a transition from olivine-dominated to pyroxene+garnet-dominated occurs around Mg/Si=1.1. High Mg/Si also tends to stabilize spinel in the upper mantle. In the lower mantle, the ferropericlase abundance increases almost linearly with Mg/Si for Mg/Si $\geq$ 0.8, where stishovite is not present. These transitions may differ somewhat with varying Ca, Al, and Na abundances, but are fairly robust (Fig.\ \ref{fig:Suppl_minscatter_MgSi}).

Both Ca and Na tend to stabilize the mineral clinopyroxene, where Ca stabilizes it at cost of garnet, while Na stabilizes it at the cost of orthopyroxene (Fig.\ \ref{fig:CaAlNa_mineralogy}). This trend is observed for a wide range of compositions within our population (Fig.\ \ref{fig:Suppl_minscatter_CaNa}). Olivine is also further stabilized by increasing Na, and to some extent by increasing Ca, as more Si is used to form clinopyroxene rather than garnet or orthopyroxene. Meanwhile, Ca only has a small effect on the Ca-pv abundance for the compositional range we find here (Fig.\ \ref{fig:Mantle_histograms}e). Both Ca/Al and Na do not significantly affect lower mantle mineralogy.

\section{Discussion} \label{sec:Disc}
\subsection{Stellar parameters} \label{ssec:Disc_Stars}
Based on stellar abundances and models for devolatilization and core-mantle partitioning, we constrain the compositional range of rocky exoplanets in the solar neighbourhood. Due to the limited range of stellar iron abundances, we find that rocky exoplanets have cores that are usually less massive than their mantles. Further, as stars tend to be rich in Fe, Mg, and Si compared to other refractory elements, we find that minor element abundances are consistently low (Table \ref{tab:comp_statistics}), and hence exotic mantle compositions are rare \citep[cf.][]{Putirka2019}. Also, the Sun and Earth compositions are close to the population median \citep{Hinkel2018,Putirka2019}. This compositional variety is mirrored in a diversity in mantle mineralogy, and thus planet properties. 

The compositional variety of stars we consider here is partly due to local variation in the solar neighbourhood, and partly due to a spread in age \citep[i.e., GCE; ][]{Burbidge1957, Frank2014, Lugaro2018}. In the GCE effect, the average heavy element content (heavier than H and He) of stars increases with time. Further, the production of iron is more efficient than that of magnesium, thereby increasing galactic Fe/Mg \citep[See Fig.\ \ref{fig:Suppl_Agetrends}a-c, \ref{fig:Suppl_Agetriangles};][]{Matteucci1986,Thielemann2002,Bensby2014}. Also, it is found that stellar Mg/Si tends to decrease with increasing Fe/H, suggesting that Mg/Si will decrease over time \citep[See Figs.\ \ref{fig:Suppl_Agetrends}d;][]{Frank2014,Adibekyan2015,Bedell2018}. Thus, newly formed planets tend to have larger cores, as we predict for HIP 38647, and mantles with larger abundances of stronger minerals such as bridgmanite, quartz, and stishovite. Finally, Ca/Al and minor element fraction only marginally evolve with time (See Figs.\ \ref{fig:Suppl_Agetrends}e,f, \ref{fig:Suppl_Agetriangles}). As stellar heavy element generally increases with time, it is more useful to compare age effects and planet properties to stellar ratios of heavy elements (e.g., Mg/Si, Fe/Mg, Ca/Al), rather than classical H-normalised compositions (e.g., cf.\ age evolution of Fe/H with Fe/Mg, Fig.\ \ref{fig:Suppl_Agetrends}a,b).

In turn, the concentration of heat-producing elements decreases over time \citep{Frank2014}, leading to less heat production in the interiors of newly formed planets. The radiogenic heat budget of a planet can contribute to thermal evolution and the propensity of a planet toward plate tectonics \citep{ONeill2007,Stein2013}. Further, \cite{Bitsch2020} find that water abundance systematically decreases with increasing metallicity, which itself decreases with age, due to the increasing abundances of C and S. This implies that planets formed around young stars are relatively dry, and should have lower oxygen fugacities than older planets, leading to even larger core mass fractions. For our selected sampled-based compositions, planets formed around the old star HIP 99651 (\textit{Sample low M$_c$}) would accrete in the most water-rich disc, while planets formed around the younger HIP 90055 (\textit{Sample high minor}) would accrete in the most water-poor disc.

Previous work has shown a difference in composition between stars of the thick and thin disc population of the Milky Way \citep{Bensby2014,Santos2017,Cabral2019}. Our Galaxy can be divided into multiple populations based on movement and age, where the latter links back to composition. For example, we find within our data that thick-disc stars are less metal-rich than thin-disc stars (Fig.\ \ref{fig:Suppl_Fe_H}). Planets in the thick-disc population also typically have higher Mg/Si and lower core mass fractions, as expected from their age (Fig.\ \ref{fig:Suppl_Agetrends}c,d, \ref{fig:Suppl_Agetriangles}). Therefore, the addition of the thick-disc population to our sample increases the width of the core size distribution. 

Eight of our sample-based representative planet compositions are based on individual stars (listed in Table \ref{tab:stellar_props}). These stars do not have any currently detected planets. The stars we chose to exemplify eight of our representative compositions show a spread of metallicity ([Fe/H]), ages, spectral types, distances, and stellar population according to our selection. Therefore, any trend of planet composition with stellar properties should, in principle, be covered by our selection. We find some minor trends between composition and spectral type (See Suppl.\ Fig.\ \ref{fig:Suppl_SpectralType}), but they are not sufficient to require splitting the population according to their host star type.

\begin{table*}[t]
\centering
\caption{Properties of the eight stars we adopt for the sample-based representative cases. Distance is from the Hypatia catalogue and is, therefore, a literature average. Ages are from various sources: [1] \cite{Stanford2020}, [2] \cite{Mints2017}, [3] \cite{DaSilva2021}, [4] \cite{Tsantaki2013}, [5] \cite{Ramirez2012}. Population refers to the distinction between the thick and thin disc stellar populations in the Milky Way \citep{Santos2017}.}\label{tab:stellar_props}
\begin{tabular}{| c | c | c | c | c | c | c | c |} 
\hline 
Planet & Star & Spectral type & Distance (pc) & Age (Gyr) & Mass ($M_{\odot}$) & [Fe/H] & Population \\
\hline
Earth & Sun & G2V & 0 & 4.5 & 1 & 0 & thin \\
\hline
Sample high M$_c$ & HIP 38647 & G3V & 36 & 1.35 [1] & 1.03 & 0.042 & thin \\
Sample low Mg/Si & HIP 83069 & F8 & 70 & 4.7 [2] & 1.15 & -0.16 & thin \\
Sample high minor & HIP 90055 & K2 & 40 & 6.47 [3] & 0.8 & 0.22 &  \\
Sample high Ca/Al & HIP 46639 & G0 & 170 & 8.38 [1] & 1.11 & -0.04 & thin \\
Sample low M$_c$ & HIP 99651 & K2V & 35 & 5.7 [4] & 0.89 & -0.80 & thin \\
Sample high Mg/Si & HIP 50493 & F6V & 36 & 2.46 [2] & 1.30 & 0.04 & thin \\
Sample high Na & HIP 51028 & G0V & 107 & 6.38 [2] & 1.13 & -0.61 & thin \\
Sample low minor & HIP 113514 & G0 & 54 & 12.45 [5] & 1.04 & -0.55 & thick \\
\hline
\end{tabular}
\end{table*}

\subsection{Devolatilization}\label{ssec:Disc_devol} 
The key link between stellar observations and our dataset of modelled terrestrial-type exoplanets is element depletion due to devolatilization. The depletion factors we use in this work are calculated according to the elemental abundances in Earth and the Sun, and the 50\% condensation temperatures of each element. Comparing various sources for Solar abundances \citep[Cf.][]{Asplund2005,Asplund2009,Lodders2009}, Earth composition \citep[Cf.][]{McDonough2003,Wang2018_Earthcomp}, and condensation temperature \citep[Cf.][]{Lodders2003,Wood2019} shows that the depletion factors we use here (see Table \ref{tab:depletion_factors}) are robust for most elements, and strongly bound by thermodynamical principles (e.g., first ionization potential of the elements). Only the depletion factor for Na varies significantly between sources (i.e., by up to 50\%), while simultaneously Na is the element with the largest depletion by devolatilization among the elements we consider here. If we had assumed Na abundances similar to bulk stellar values in Figure \ref{fig:CaAlNa_mineralogy}, the corresponding range would be 0.4 to 1.4 mol\% Na$_2$O. Decreasing the Na depletion factor would have little effect on lower mantle mineralogy, but would further stabilize clinopyroxene, and may lead to even lighter and more buoyant crusts (cf.\ Earth and Sun in Figure \ref{fig:CaAlNa_mineralogy}).

The devolatilization trend shows that planets are progressively more depleted in elements with lower condensation temperatures, below a certain cut-off temperature \citep[around 1400 K][]{Wang2019}. The slope of this depletion trend depends mainly on the width of the feeding zone \citep[cf.\ Earth to Vesta;][]{Sossi2022}, while the cut-off temperature depends mainly on distance to the host star. Applying the Earth-Sun devolatilization trend to exoplanets inherently assumes that the planet is formed in the habitable zone of its host star, and its feeding zone samples material with a similar range of volatile depletion as Earths. Planet migration and disc dynamics may affect the cut-off temperature and slope of the devolatilization disc. Within the elements we consider, mainly affecting the depletion of Na, which would be more abundant for planets forming further out. The moderately refractory elements Fe, Mg, and Si could be more depleted in planets forming closer to the star than considered here, leading to an enrichment in Ca and Al.

Depletion factors are based on condensation temperatures for multiple composite minerals, which for most of our elements are similar \citep{Lodders2003}. Hence, moderate compositional variation will not significantly affect 50\% condensation temperatures, perhaps except for extreme stellar compositions \citep[e.g., sulphur for extremely low Fe/S; ][]{Jorge2022}, which are, however, beyond the range of our data. How composition will affect planetary devolatilization is still an ongoing investigation \citep{Wang2020_EPSC,Sossi2022_EPSC}. For example, the condensation temperature of oxygen varies with stellar composition due to its binary nature as a refractory element (in silicates) and a volatile element (in volatiles such as water). Within our assumption of constant \textit{fO$_2$}, our planet oxygen budgets imply that 6.8 to 28\% of available oxygen (see Suppl.\ Fig.\ \ref{fig:Suppl_RO}) has condensed as refractory compounds. The corresponding effective condensation temperatures for oxygen range from 721 to 968K, comparable to the value for Earth from \cite{Wang2019} of 875 $\pm$ 45K.

Planet formation is complex and chaotic \citep{Morbidelli2016}, but the stochastic nature of this process yields a smooth pattern of volatile depletion for planets like Earth, dictated by the central limit theorem, despite various thermal and non-thermal effects \citep{Sossi2022}. This supports that an Earth-like devolatilization trend is a general result of rocky planet formation. For example, \cite{Wang2022_planethoststars} argue that the devolatilization trends of Venus and Mars may not be significantly different from that of Earth, based on the currently yet-large uncertainties in their individual bulk compositions \citep[e.g.,][]{Morgan1980, Taylor2013, Sossi2018, Wang2018_Earthcomp, Yoshizaki2020,Shah2022}. If we arbitrarily increase the uncertainty range of the adopted Sun-to-Earth devolatilization model, as practised in \cite{Wang2022_planethoststars}, it would increase the uncertainties of our compositional estimates for individual planets. However, the results of both our population analysis and their bounds in terms of our representative compositions should remain robust, particularly in terms of major elements, which are almost unaffected by devolatilization (Fig.\ \ref{fig:Bulk_triangles}a). 

Our approach excludes a few classes of extreme planet compositions. Firstly, extremely C-rich planets have been theorized for stars with molar C/O greater than 0.8 \citep{Bond2010,Moriarty2014}, suggested to form a thick crust of graphite and diamond \citep{Hakim2019}. We find that only about 5\% of the stars in the Hypatia and GALAH catalogues have C/O ratios greater than 0.8, most of which have error bars stretching below 0.8. Secondly, Ca-Al-rich planets are theorized to form in environments close to the star, where most elements except Ca and Al cannot condense due to the high temperature \citep{Dorn2019}. These planets are implicitly assumed to form in situ without having accreted mixed material likely transported from the outer to the inner disc. 
However, due to the low abundances of Ca and Al compared to Fe, Mg, Si, these planets (if they exist) will mostly be very small. Lastly, water-worlds have been predicted, i.e., terrestrial planets with a thick layer of water and ice on the surface \citep[e.g.][]{Kuchner2003,Unterborn2018, Acuna2021,Krissansen-Totton2022}. 
If such a layer of water is sufficiently large, it could fundamentally alter planetary interior dynamics by suppressing mantle melting and crust formation \citep{Unterborn2018}. However, water delivery to rocky planets is a highly debated topic, and it is currently not possible to link water delivery to stellar observations.

\subsection{Core-mantle differentiation} \label{ssec:Disc_coremantle}
We find core mass fractions (CMF) ranging from 18 to 35 wt\% (Fig.\ \ref{fig:Core_masses}), indicating that Earth (32.3 wt\%) has a relatively large core compared to the modelled exoplanet population. We underestimate the Earth CMF based on Solar abundances by 1.5 wt\% (Fig.\ \ref{fig:Core_masses}). The Earth has elevated Fe/Mg compared to the Sun, which could be attributed to secondary fractionation processes \citep{ONeill2008}, or it can be treated as a statistical residual to the devolatilization trend \citep[see ][ for more details]{Wang2022_planethoststars}. Overall, we underestimate exoplanet CMF by about 1.5 wt\%, which is within measurement error. Core size correlates well with stellar [Fe/Mg] (Fig.\ \ref{fig:Core_masses}), better than with [Fe/H] \citep[e.g.,][]{Hinkel2018}. Modelled interior structures based on observed terrestrial-type exoplanet masses and radii display a fairly similar range of CMF \citep[typically 20-41 wt\%;][]{Otegi2020,Plotnykov2020,Adibekyan2021,Schulze2021}, except for a small population of super-Mercuries, which have significantly higher CMFs, potentially due to secondary processes \citep[e.g.,][]{Aguichine2020,Scora2020,Adibekyan2021}. A large portion of the planets in our dataset have a smaller core than Earth and thus are expected to maintain hotter interiors \citep{Noack2014_structure}, and are more likely to develop plate tectonics \citep{ONeill2020_GCE}. Core size is only marginally affected by the presence of Ni and S, as we find core light element content varying by only a few per cent (\ref{fig:Suppl_Core_comp}a). Molar Fe/Ni can be up to 30\% lower than on Earth, leading to slightly larger cores with higher Ni content (\ref{fig:Suppl_Core_comp}c,d). However, the resulting increase in core size is less than one per cent, and there is no correlation between core Ni content and core mass fraction (\ref{fig:Suppl_Core_comp}d).

In our study, we base the core-mantle differentiation process on the assumption that our modelled planets have oxygen fugacities similar to Earth. Allowing \textit{fO$_2$} to vary could potentially result in planets with core sizes up to 45 wt\% (all iron in the core, given our assumed core light element compositions), or could, in turn, result in coreless exoplanets \citep{Elkins2008_coreless, Wang2019MNRAS}. The oxidation state of a terrestrial planet interior depends mainly on the accretion rate and timing of influx from planetesimals from further out in the planet-forming disc \citep{Monteux2018}, which tend to be more oxidized \citep{Rubie2015,Monteux2018,Cartier2019}. Oxygen fugacity should vary only moderately (even though to an unknown extent) between planets that orbit their host star in the habitable zone and with masses similar to Earth, as we assume here. Nevertheless, we consider representative compositions with different fO$_2$, and explore the effect of varying fO$_2$ on the distribution of iron between mantle and core (Fig.\ \ref{fig:Suppl_Fe_FeO}).

\subsection{Compositional effects} \label{ssec:Disc_comp}
Based on the inferred variation of terrestrial-type exoplanet compositions, we identify 21 representative planet compositions (\ref{tab:endmember_comps}). These representative compositions result in mantle mineral profiles that are usually very similar to, or even the same as, Earth. However, the relative abundances of inferred mineral species vary significantly between planets, potentially influencing interior properties and, thus, long-term evolution. Thermal evolution of the interior is controlled by convective transport of heat from the core-mantle boundary to the surface. Convective vigour of mantle material increases with increasing thermal Rayleigh number \citep[e.g.,][]{Schubert2001}, which is given by
\begin{equation}
    Ra = \frac{g \rho \alpha \Delta T d^3}{\kappa \eta},
    \label{eq:Rayleigh}
\end{equation}
for gravitational acceleration $g$ (m s$^{-2}$), density $\rho$ (kg m$^{-3}$), thermal expansivity $\alpha$ (K$^{-1}$), temperature contrast across the mantle $\Delta T$ (K), mantle thickness $d$ (m), thermal diffusivity $\kappa$ (m$^2$ s$^{-1}$), and viscosity $\eta$ (Pa s). Of these parameters, the compositional effect on $\alpha$ and $\beta$ is less significant than on $d$ through core size, and on $\eta$ through mineralogy.

\subsubsection{Mantle viscosity}
Bulk mantle viscosity is regulated by relative abundances of strong and weak mineral phases (i.e., high- and low-viscosity phases). In the lower mantle, bridgmanite is up to three orders of magnitude stronger than ferropericlase \citep{Yamazaki2001,Tsujino2022}. While all planets have lower mantles with high bridgmanite abundances, some are almost completely lacking in ferropericlase (Fig.\ \ref{fig:Mineralogy}c), while others have significant abundances of the weak mineral ferropericlase (Fig.\ \ref{fig:Mineralogy}b). These planets are expected to have a strongly contrasting lower-mantle viscosity profile, as the weaker phase tends to have a more significant effect on viscosity due to formation of interconnected weak layers \citep{Yamazaki2001,Thielmann2020}. Further, our representative planets exhibit variable but overall small amounts of Ca-perovskite. The phase Ca-pv has typically been considered to be even stronger than bridgmanite and ferropericlase \citep{Miyagi2009}, but recent experimental studies indicate that it may be significantly weaker than both these minerals \citep{Shieh2004,Immoor2022}. The highest Ca-pv abundance of our compositions is 10 vol\%, which may be sufficient to form interconnected weak layers \citep{Yamazaki2001,Thielmann2020}, thereby decreasing mantle viscosity, if it is indeed weaker than other lower mantle phases. 

Meanwhile, the upper mantle viscosity is mainly controlled by olivine and pyroxene. In the upper mantle, our representative planet compositions show a huge range from 10\% to 80\% olivine (Fig.\ \ref{fig:Mineralogy}). While surface observations indicate that pyroxene is stronger than olivine \citep{Tikoff2010}, other studies suggest that the px/ol-ratio has limited effects on bulk rock viscosity as long as both phases are present \citep[e.g.][]{Tasaka2013,Hansen2015}. Some of our representative planets exhibit ferropericlase in the upper mantle, a phase with a viscosity about one order of magnitude lower than that of olivine or pyroxene \citep{Stretton2001,Bystricky2006}. Observations of polluted white dwarfs in the Solar neighbourhood confirm the possibility of an upper mantle containing ferropericlase \citep{Putirka2021}. While the presence of ferropericlase in exoplanet upper mantles has been derived by previous studies \citep[e.g.,][]{Wang2022_planethoststars}, it is currently not well studied. It is likely to be relevant for planet evolution studies, as it directly affects upper mantle viscosity, which is an important parameter for determining the propensity of a planet towards plate tectonics \citep{Korenaga2010,vanHeck2011}.

The polluted white dwarf observations also indicate the presence of planets with quartz in the upper mantle, and \cite{Putirka2019} find a tiny population of planets where pure SiO$_2$ could potentially be present in their mantles (although they do not employ a Gibbs energy minimisation algorithm). While these compositions fall outside our range of representative compositions, they are permissive considering measurement errors (Fig.\ \ref{fig:MgSi_mineralogy}). Further, assuming that the core does not contain Si decreases the mantle Mg/Si of our lowest-Mg/Si composition such that a few vol\% of quartz, coesite, and stishovite are stabilised. We expect the abundance of SiO$_2$-phases in these planets to be small, however (See Fig.\ \ref{fig:MgSi_mineralogy}), and only have a minor impact on planetary evolution.

We predict mantle mineralogy of planets with our representative compositions (Tab.\ \ref{tab:endmember_comps}), using the Gibbs energy minimization algorithm Perple\_X \citep{Connolly2005}. To accurately determine planet mantle mineralogy, the stellar composition should be known to a precision of $\lesssim$ 0.025 dex \citep{Wang2022_planethoststars}, which is significantly smaller than available uncertainties in the Hypatia and GALAH catalogues, and the vast majority of our planets cannot be constrained to this level of accuracy. That is why we approach this task by considering the entire population; while individual planet compositions should be regarded with appropriate caution, the overall trends we find here are robust. Further, our 21 sets of representative planets are meant to illustrate the limits of the population in terms of composition, which will not change significantly even when individual planet compositions shift within the current abundance uncertainties.

\subsubsection{Other compositional effects}
Melting behaviour of planetary mantles is influenced by both the mineral species in the upper mantle (where the vast majority of melting occurs) and the iron content of those minerals. Olivine melts at higher temperatures than pyroxene and garnet, and iron-rich olivines and pyroxenes melt at lower temperatures than their Fe-poor equivalents \citep{Hirschmann2000,Kiefer2015}. Planet mantles with higher Ca/Al have also been found to melt at lower temperatures than their low-Ca/Al counterparts, even when mineralogy is otherwise very similar \citep{Brugman2021}. Therefore, we expect some variation in the degree of melting and volcanism among our planets, which will in turn affect mantle outgassing and atmosphere-interior interaction \citep[e.g.,][]{Noack2017,Dorn2018,Spaargaren2020,Gaillard2021}. Further, due to the thermostat effects caused by temperature-dependence of viscosity \citep{Tozer1965} and, for hot planets, by magmatism \citep{Ogawa2011}, the mantle geotherm is expected to evolve close to the solidus of the upper mantle. Thus, the upper mantle mineralogy strongly affects planetary thermal evolution. 

Aside from viscosity, mineralogy also affects mantle dynamics in different ways. On Earth, the Ringwoodite-Bridgmanite+Ferropericlase phase transition at 660 km depth creates a boundary between the upper and lower mantle that impedes convection \citep{Schubert1975,Christensen1985}. Some of our planets are richer in Ringwoodite than Earth, which could perhaps lead to doubled-layered convection instead of whole-mantle convection. This is most likely to happen where the olivine fraction is highest, around Mg/Si=1.5. In contrast, some of our planets have very little Ringwoodite, potentially promoting efficient material exchange between the upper and lower mantle (i.e., whole-mantle convection). Further, water storage capacity in the mantle is expected to vary strongly between exoplanets. Minerals with high water storage capacity, such as wadsleyite and ringwoodite \citep[e.g.,][]{Kohlstedt1996}, are rare in low-Mg/Si planets, which would then be expected to have most of their water budget partitioned to the surface. This difference will impact mantle dynamics \citep{Hirth1996,Korenaga2011}, melting \citep{Katz2003}, and volatile outgassing \citep{Wang2022_alphacentauri}. As both Na and Ca tend to stabilize olivine and its high-pressure polymorphs, these elements contribute towards increased water storage capacity in the mantle, and to potential for double-layered convection. Finally, we expect crustal composition to vary as a function of bulk planet composition. Planets rich in Na and Si will have more buoyant crust than Earth, which may render subduction and hence plate tectonics less efficient \citep{Cloos1993,Unterborn2017}. 

We described earlier that stellar composition evolves over time, with recently formed star being richer in Fe and having lower Mg/Si than stars formed long ago (Sect.\ \ref{ssec:Disc_Stars}). Most of these age-related effects work towards changing the typical thermal evolution pathway of a terrestrial planet. The Rayleigh number (eq.\ \ref{eq:Rayleigh}) decreases when the material becomes more viscous, either due to lower Mg/Si forming stronger minerals, or due to a lower water fraction \citep[e.g.,][]{Chopra1984}. Further, a larger core makes for a smaller mantle, which decreases the Rayleigh number further. Finally, a higher radiogenic element budget increases internal heating, which in turn increases convective vigour. These factors affect the propensity towards plate tectonics, but are often associated with conflicting results \citep[likelihood as a function of Rayleigh number, cf.][]{Korenaga2010,vanHeck2011}. How these effects compare, and what the effect of time of formation is on thermal evolution of a planet and the propensity towards plate tectonics, requires investigation with geodynamical models and will be the subject of future research.

\section{Conclusions} \label{sec:Conc}
We present the plausible range of bulk terrestrial exoplanet compositions in the solar neighbourhood by considering terrestrial planets as devolatilized stars compositionally. This approach is based on the assumption that planets form from materials condensed from a nebula that shares the chemical composition of the host star. Further, assuming (to the first order) that devolatilization is a universal process in forming rocky planets, we applied the Earth-Sun depletion factors to a large set of stellar abundances in the Solar neighbourhood (within 200 pc) and obtain the plausible range as presented. 

We find that the compositions of the Sun and Earth are close to the medians of the bulk compositions of the population of Sun-like stars and the postulated terrestrial-type exoplanets around them. For example, rocky exoplanets span a wide range of relative abundances of Mg and Si, with the Earth and Sun close to averages. Further, exotic compositions (i.e., compositions deviating significantly from Earth's) are rather rare, with most planets having mantles of 90-95 molar\% MgO+FeO+SiO$_2$. Core sizes range from 18 to 35 wt\%, and show a strong correlation with stellar [Fe/Mg].

Stellar Mg/Si is a valuable indicator for mantle mineralogy. To first order, it can be used to predict lower mantle mineralogy, specifically the ratio of the strong mineral bridgmanite to the weak mineral ferropericlase, and thus lower mantle viscosity. In the upper mantle, it can indicate the presence of ferropericlase, which has yet to be well studied.
The mantle Mg/Si can indicate a transition from ferropericlase-bearing upper mantles (mantle Mg/Si $\geq$ 1.6), to strong, quartz-bearing (mantle Mg/Si $\leq$ 0.8). Further, both Fe/Mg, which indicates core size, and Mg/Si are correlated with age. Recently formed planets tend to have large cores and stiff mantles, while planets formed long ago tend to have small cores and weak mantles. Additionally, stellar Mg/Si and Na/Mg can be used as indicators for crustal bouyancy, as crusts of Na- and Si-rich planets tend to have lower density than crusts of Na- and Si-poor planets. Finally, stars rich in Ca and Fe lead to planet mantles with lower melting temperatures, and can therefore lead to increased volcanism. Thus, stellar composition, upon a necessary correction of devolatilization, gives a more accurate and complete picture of planet properties. Importantly, omitting the correction on stellar abundances would reveal a different nature towards the crustal buoyancy. These properties all contribute towards a planets tendency towards a mobile lid regime or a stagnant lid regime, but much research remains to be done before we can couple stellar composition to this tendency.

Rocky planet composition affects planetary evolution in multiple ways, and it is therefore a crucial component in understanding how rocky planets work. We have taken a first step towards studying this component by placing constraints on the range of potential bulk compositions of terrestrial-type exoplanets. By considering 21 representative compositions that span the full compositional range, we also deliver a convenient reference dataset for further numerical, experimental and observational studies on the effects of rocky exoplanet compositions in the broad context of exoplanet characterization. 
\\
\\
\\
RJS has been funded by ETH grant number ETH-18 18-2. Contributions of HSW have been carried out within the framework of the NCCR PlanetS supported by the Swiss National Science Foundation under grants 51NF40\_182901 and 51NF40\_205606. SJM\ extends a special thanks to C.\ Heubeck at the Institute for Earth Sciences at the Friedrich-Schiller University of Jena (Germany) and the A.v.\ Humboldt Foundation that provided support during the final writing of this manuscript. SJM\ is supported by the Research Centre for Astronomy and Earth Sciences in Budapest (Hungary). The research shown here acknowledges use of the Hypatia Catalog Database, an online compilation of stellar abundance data as described in \cite{Hypatia}, which was supported by NASA's Nexus for Exoplanet System Science (NExSS) research coordination network and the Vanderbilt Initiative in Data-Intensive Astrophysics (VIDA). This work also made use of the Second Data Release of the GALAH Survey \citep{Buder2018}. The GALAH Survey is based on data acquired through the Australian Astronomical Observatory, under programs: A/2013B/13 (The GALAH pilot survey); A/2014A/25, A/2015A/19, A2017A/18 (The GALAH survey phase 1); A2018A/18 (Open clusters with HERMES); A2019A/1 (Hierarchical star formation in Ori OB1); A2019A/15 (The GALAH survey phase 2); A/2015B/19, A/2016A/22, A/2016B/10, A/2017B/16, A/2018B/15 (The HERMES-TESS program); and A/2015A/3, A/2015B/1, A/2015B/19, A/2016A/22, A/2016B/12, A/2017A/14 (The HERMES K2-follow-up program). We acknowledge the traditional owners of the land on which the AAT stands, the Gamilaraay people, and pay our respects to elders past and present. This paper includes data that has been provided by AAO Data Central (datacentral.org.au).

\bibliographystyle{aasjournal}
\bibliography{Mendeley}

\section*{Supplementary material}
\renewcommand{\thefigure}{S\arabic{figure}}
\setcounter{figure}{0}

\subsection*{Data conversion and error propagation}
Stellar abundance data in the Hypatia catalogue are available in \textit{dex} units, [X/H] for element X, while data in the GALAH catalogue are available as [X/Fe], alongside [Fe/H]. To convert these to bulk planet molar composition for the relevant elements (O, Na, Al, Mg, Si, S, Ca, Fe, Ni), we adopt an approach similar to \cite{Hinkel2022}, 
with the addition of applying depletion factors from Table \ref{tab:depletion_factors}, which we label $d_i$, where $i = 1 \dots 9$ corresponds to the 9 elements we consider here. We first retrieve bulk planet abundances $X_i$ before normalizing to bulk planet concentration in molar ppm $C_i$. Similar to equation 14 of \cite{Hinkel2022}, and using solar abundances $X_{\odot}$ from \cite{Lodders2009}, we find bulk planet abundances
\begin{equation*}
    X_i = 10^{[X_i/H] + X_{\odot,i} + d_i},
\end{equation*}
for the Hypatia data and for GALAH Fe abundances, while for other GALAH abundances, given in [X/Fe], we use
\begin{equation*}
    X_i = 10^{[X_i/Fe] + [Fe/H] + X_{\odot,i} + d_i}.
\end{equation*}
Finally, we sum up the abundances of all 9 elements we consider here, and normalize to retrieve the bulk planet composition in molar ppm, $C_i = 10^6 * X_i / \sum_{i=1}^9 X_i$.

Regarding error propagation, we adopt an approach similar to equations 19 and 20 from \cite{Hinkel2022}, which we adapt for an increased amount of variables. Following \cite{Taylor1997}, we know that error propagation of a function $f(x,y)$ follows $\sigma_f = \sqrt{\left(\frac{\partial f}{\partial x}\right)^2\sigma_x^2 + \left(\frac{\partial f}{\partial y}\right)^2\sigma_y^2}$. Therefore, for abundances from the Hypatia catalog and Fe abundances from the GALAH catalog, we have
\begin{equation*}
    \sigma_{X_i} = \ln(10) \sqrt{\sigma_{[X_i/H]}^2 + \sigma_{d_i}^2},
\end{equation*}
where $\sigma_{[X_i/H]}$ is given by the respective catalogue, and $\sigma_{d_i}$ is given by Table \ref{tab:depletion_factors}. Typical stellar errors are between 0.02 and 0.10 dex. For other elements in the GALAH catalogue, we have
\begin{equation*}
    \sigma_{X_i} = \ln(10) \sqrt{\sigma_{[X_i/Fe]}^2 + \sigma_{[Fe/H]}^2 + \sigma_{d_i}^2},
\end{equation*}
where $\sigma_{[X_i/Fe]}$ are given by the catalogue. Finally, for converting $\sigma_{X_i}$ to $\sigma_{C_i}$, we need to apply error propagation again, giving us
\begin{equation*}
 \sigma_{C_i} = \frac{C_i}{\sum_{j=1}^9 X_j} \sqrt{\left( \sum_{j \neq i} X_j / X_i \right)^2\sigma_{X_i}^2 + \sum_{j \neq i}\sigma_{X_j}^2}.
\end{equation*}

\subsection*{Age trend estimates}

\begin{figure*}[h]
    \resizebox{\hsize}{!}{\includegraphics{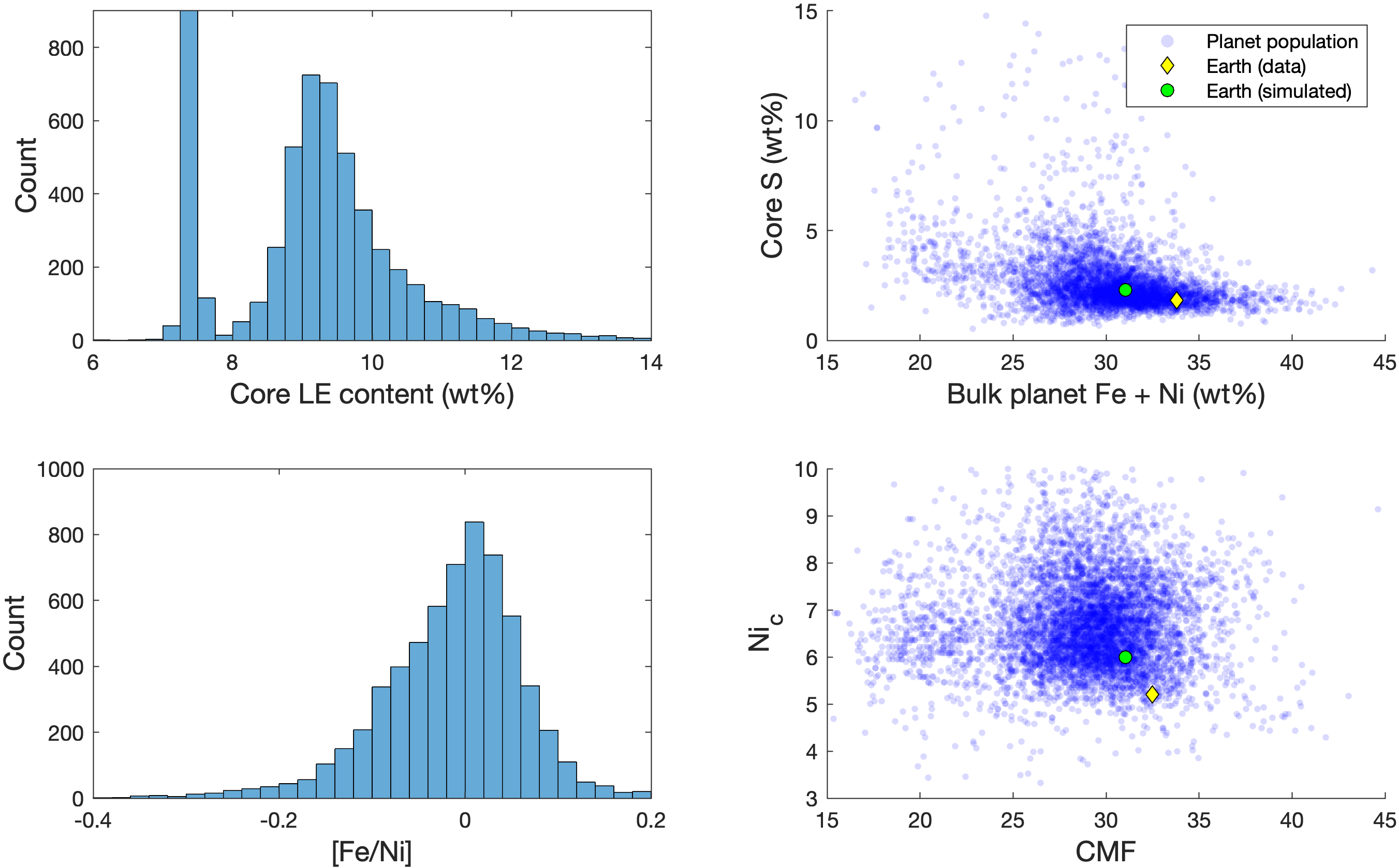}}
      \caption{Composition of the cores of exoplanets, compared to Earth from \cite{McDonough2003}, and a simulated Earth based on the Solar composition from ... . We plot core light element content in wt\%, which consists of O, Si, and S (a). The large peak at 7.5 wt\% is caused by planet compositions based on stellar compositions from the GALAH catalogue, which does not contain sulphur abundances. Further, core S concentration in wt\% is plotted against bulk planet Fe+Ni content (b). We plot the distribution of stellar [Fe/Ni] (c). Finally, we plot core molar Ni content against bulk core mass fraction (d).}
      \label{fig:Suppl_Core_comp}
\end{figure*}

\begin{figure*}[h]
    \resizebox{\hsize}{!}{\includegraphics{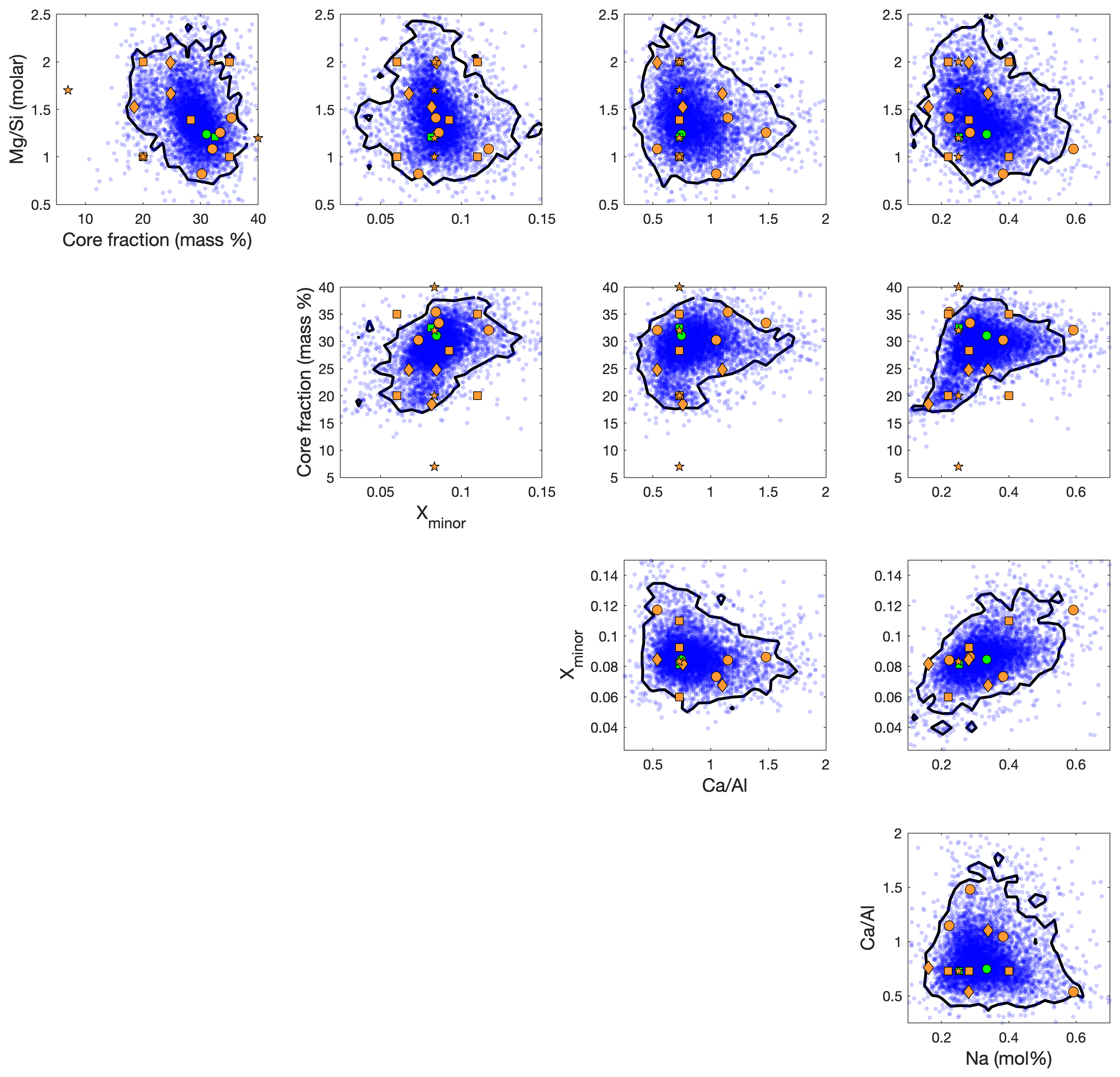}}
      \caption{Scatter plots of bulk terrestrial exoplanet compositions for all compositional quantities, showing compositional trends. The contours contain 95\% of the compositions. Orange dots and diamonds show the high-Fe/Mg and low-Fe/Mg \textit{sample-based} compositions, orange squares show the eight \textit{synthetic} compositions, respectively, and the stars show the four \textit{synthetic-fO$_2$} compositions. The green square represents Earth composition \citep{McDonough2003}, while the green circle represents the composition of a simulated planet based on a star with a protosolar composition \citep{Wang2019}.}
      \label{fig:Suppl_scatter}
\end{figure*}

\begin{figure*}[h]
    \resizebox{\hsize}{!}{\includegraphics{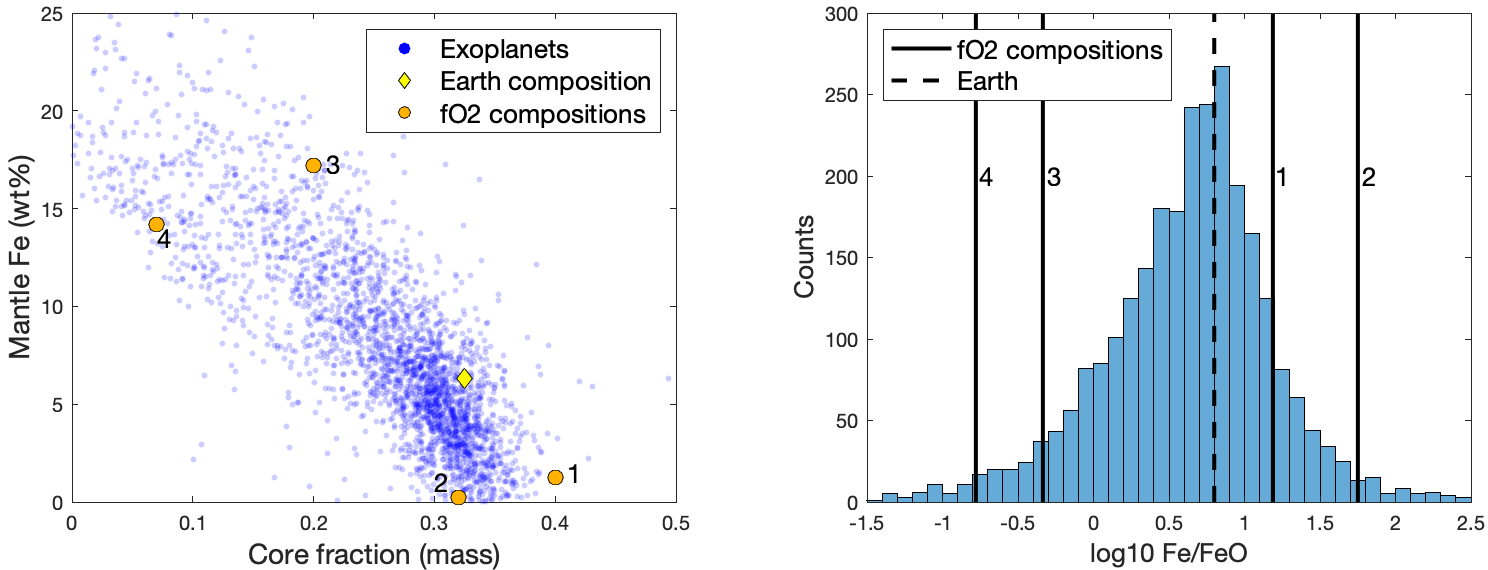}}
      \caption{Scatterplot of core mass fraction and mantle iron content (left), and histogram of corresponding bulk planet Fe/FeO ratio (right), without the assumption of constant \textit{fO$_2$} for all planets (see Sect.\ \ref{ssec:Res_endmembers}). Bulk oxygen fugacity is based on stellar composition after devolatilization, after adding a correction for removing coreless planets and extremely reduced planets. Representative compositions \textit{synthetic-fO$_2$} 1-4 from Table \ref{tab:endmember_comps} are plotted in both figures.}
      \label{fig:Suppl_Fe_FeO}
\end{figure*}

\begin{figure*}[h]
    \resizebox{\hsize}{!}{\includegraphics{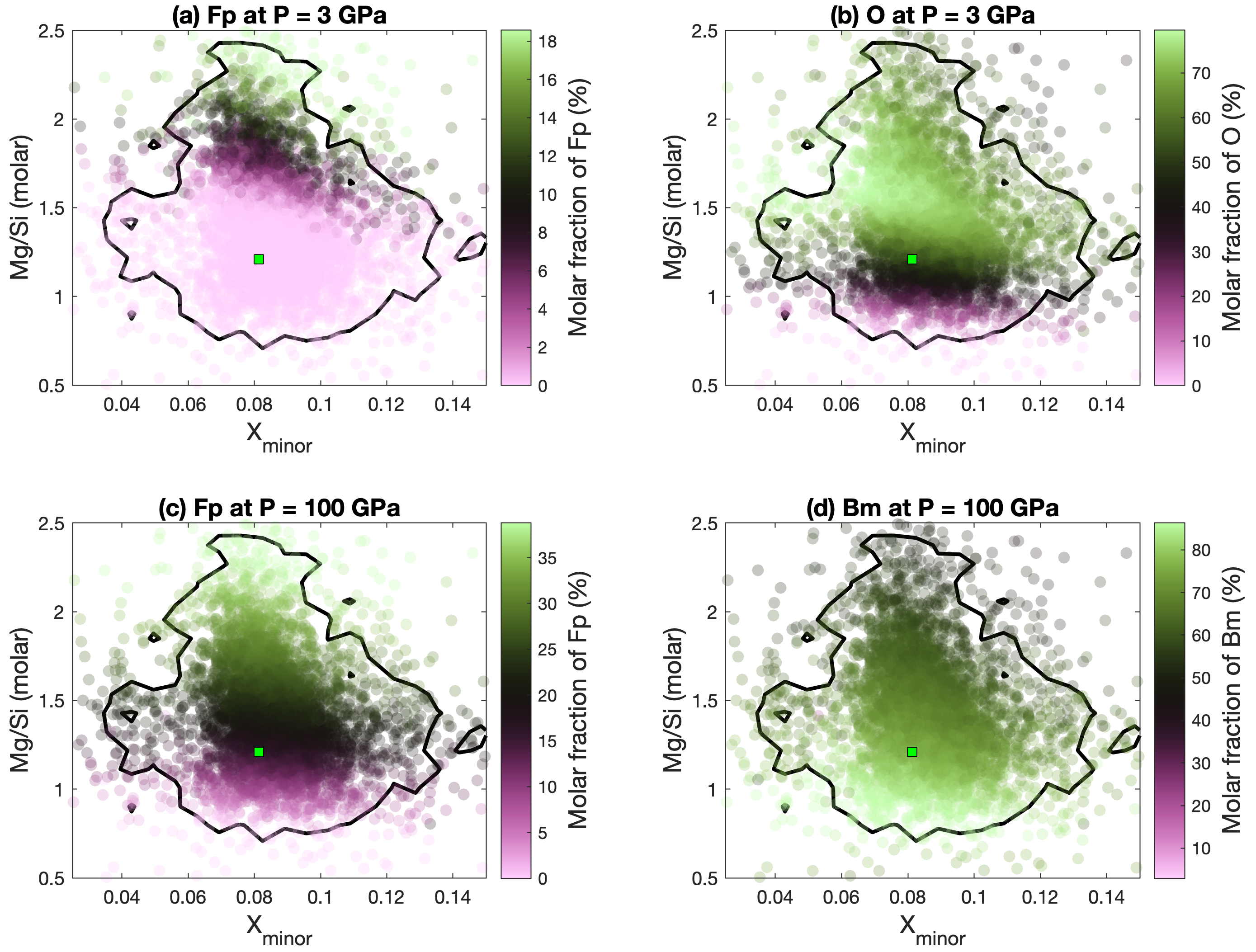}}
      \caption{Molar fractions of ferropericlase (a,c), olivine (b), and bridgmanite (d), at 3 (top) and 100 (bottom) GPa. Mineralogical assemblages are calculated from mantle compositions based on stellar abundances for all 6207 datapoints in our population, using Perple\_X \citep{Connolly2005} with the thermodynamic database from \cite{Stixrude2022}. Earth composition (yellow square) is from \cite{McDonough2003}.}
      \label{fig:Suppl_minscatter_MgSi}
\end{figure*}

\begin{figure*}[h]
    \resizebox{\hsize}{!}{\includegraphics{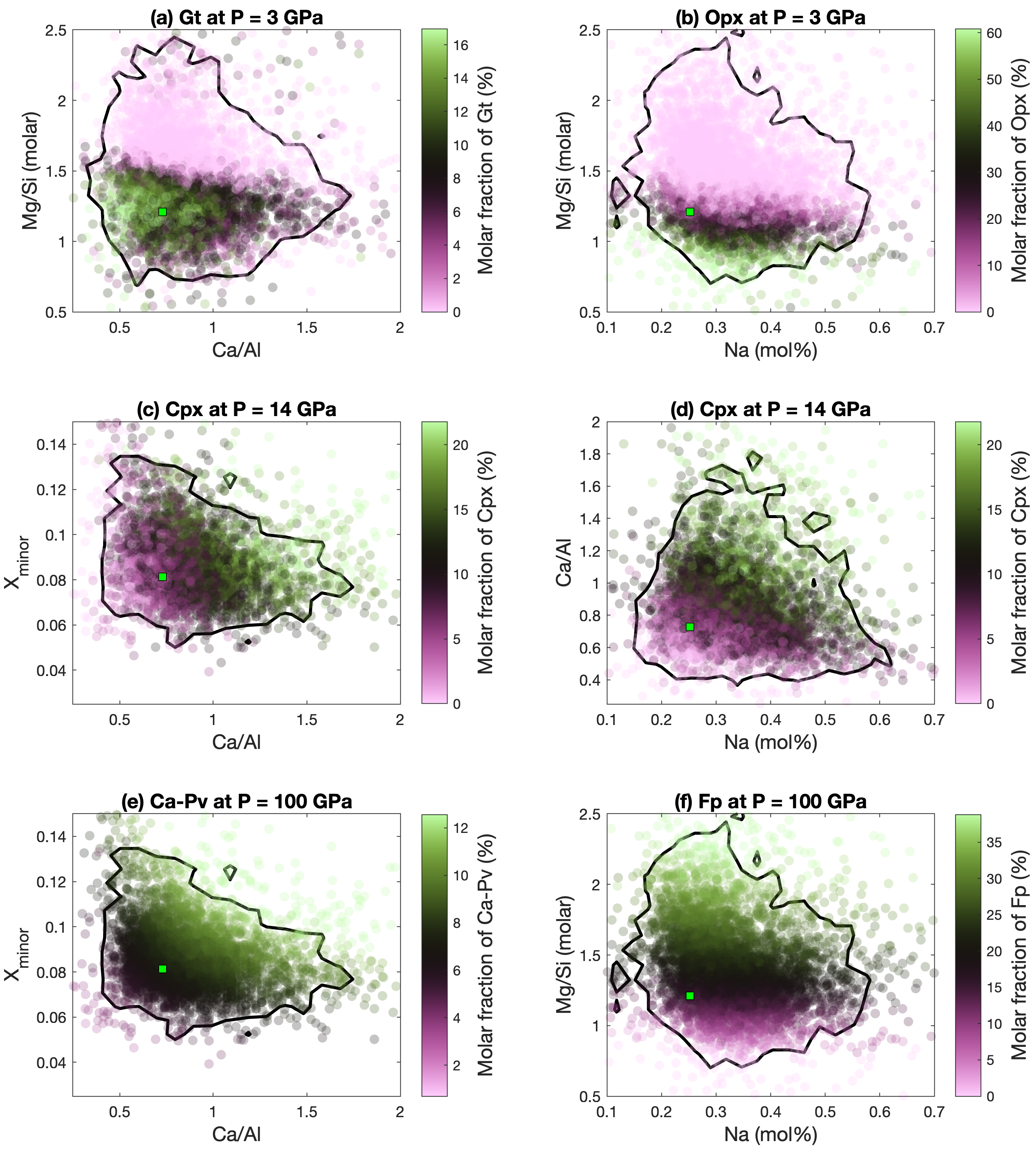}}
      \caption{Molar fractions of garnet (a), orthopyroxene (b), clinopyroxene (c,d), Ca-perovskite (e), and ferropericlase (f), at 3 (top), 14 (middle), and 100 (bottom) GPa. Mineralogical assemblages are calculated from mantle compositions based on stellar abundances for all 6207 datapoints in our population, using Perple\_X \citep{Connolly2005} with the thermodynamic database from \cite{Stixrude2022}. Earth composition (yellow square) is from \cite{McDonough2003}.}
      \label{fig:Suppl_minscatter_CaNa}
\end{figure*}

\begin{figure*}[h]
    \resizebox{\hsize}{!}{\includegraphics{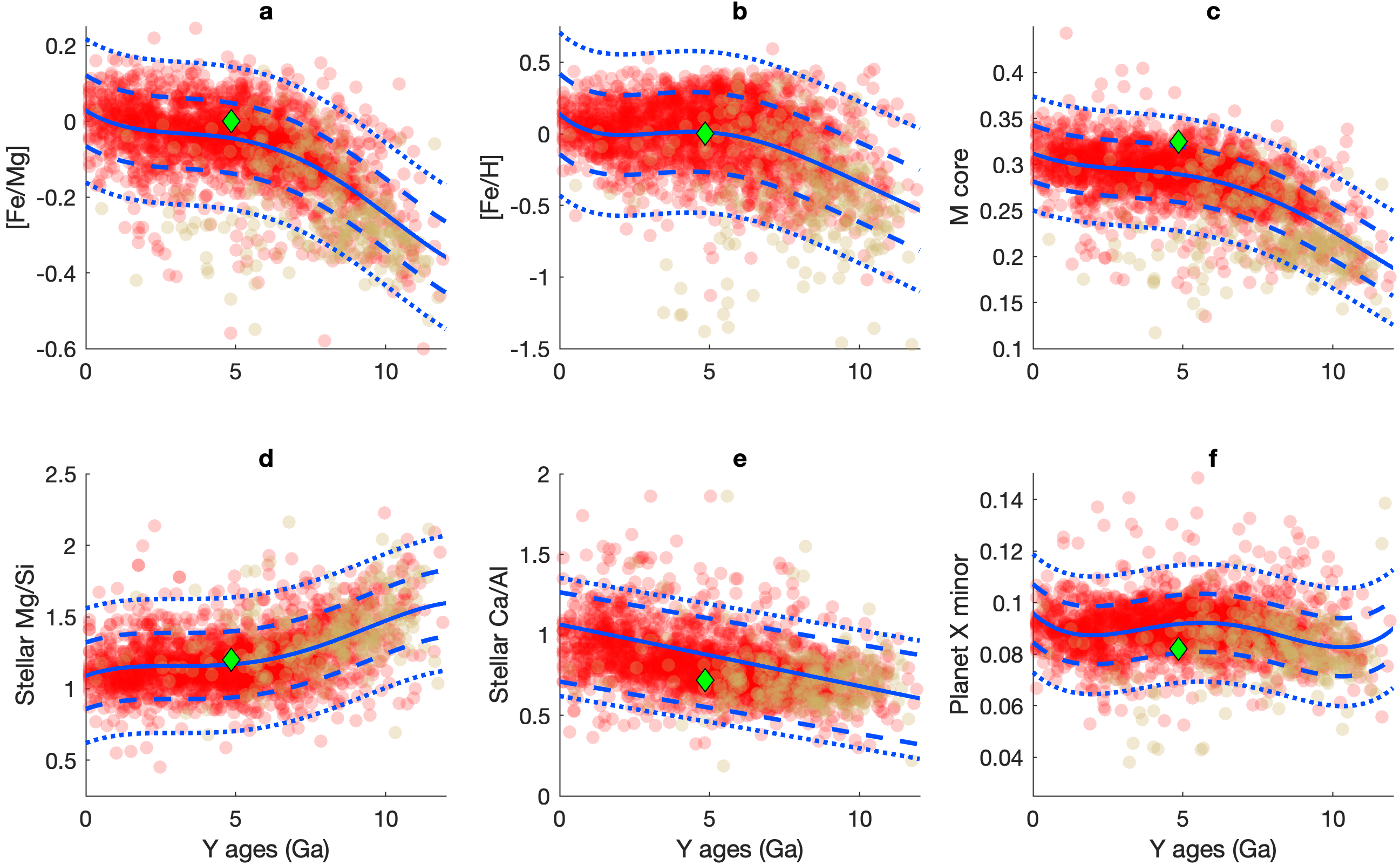}}
      \caption{Stellar [Fe/Mg] (a), stellar [Fe/H] (b), core mass fraction (c), stellar molar Mg/Si (d), stellar Ca/Al (e), and planet mantle minor element fraction (f) as a function of stellar age (in Ga), estimated as a function of Y/Mg and Y/Al, based on equations 6 and 7 from \cite{Spina2018}. Stellar compositions are from \cite{Hypatia}, colour-coded for the thin disc (red) and thick disc (gold) populations of the Milky Way.}
      \label{fig:Suppl_Agetrends}
\end{figure*}

We fit polynomials to the age-composition trends shown in Figure \ref{fig:Suppl_Agetrends}. The order of the polynomial is decided based on when the RMSE of the fit stops increasing significantly when adding terms, while the p-values of each term are below 0.05. These fits do not change significantly when only considering thin-disc stars. The six polynomials are given as a function of stellar age $a$, based on the average of Y/Mg- and Y/Al-ages from Equations 6 and 7 from \cite{Spina2018}.
\begin{equation}
     \text{M}_{\text{core}}  = 0.3122 \pm 0.0069 - 0.0124 \pm 0.00624 a + 3.53 \pm 1.75 a^2 \cdot 10^{-3} - 4.89 \pm 1.85 \cdot 10^{-4} a^3 + 1.72 \pm 0.65 \cdot 10^{-5} a^4,
\end{equation}
\begin{equation}
    [\text{Fe/H}] = 0.140 \pm 0.077 - 0.185 \pm 0.066 a + 0.079 \pm 0.040 a^2 - 0.0135 \pm 0.0069 a^3 + 9.34 \pm 5.20 \cdot 10^{-4} a^4 - 2.33 \pm 1.39 \cdot 10^{-5} a^5,
\end{equation}
\begin{equation}
    [\text{Fe/Mg}] = -0.062 \pm 0.021 - 0.049 \pm 0.019 a + 0.0152 \pm 0.0054 a^2 - 2.07 \pm 0.57 \cdot 10^{-3} a^3 + 7.68 \pm 1.99 \cdot 10^{-5} a^4,
\end{equation}
\begin{equation}
    \text{Mg/Si} = 1.089 \pm 0.052 + 0.073 \pm 0.047 a - 0.027 \pm 0.013 a^2 + 3.97 \pm 1.41 \cdot 10^{-3} a^3 - 1.6 \pm 0.5 \cdot 10^{-4} a^4,
\end{equation}
\begin{equation}
    \text{Ca/Al} = 0.987 \pm 0.018 - 0.0325 \pm 0.0032 a,
\end{equation}
\begin{equation}
    \text{X}_{\text{minor}} = 0.0959 \pm 0.0025 - 0.0105 \pm 0.0023 a + 4.10 \pm 0.65 \cdot 10^{-3} a^2 - 5.47 \pm 0.67 \cdot 10^{-4} a^3 + 2.29 \pm 0.24 \cdot 10^{-5} a^4,
\end{equation}

\begin{figure*}[h]
    \resizebox{\hsize}{!}{\includegraphics{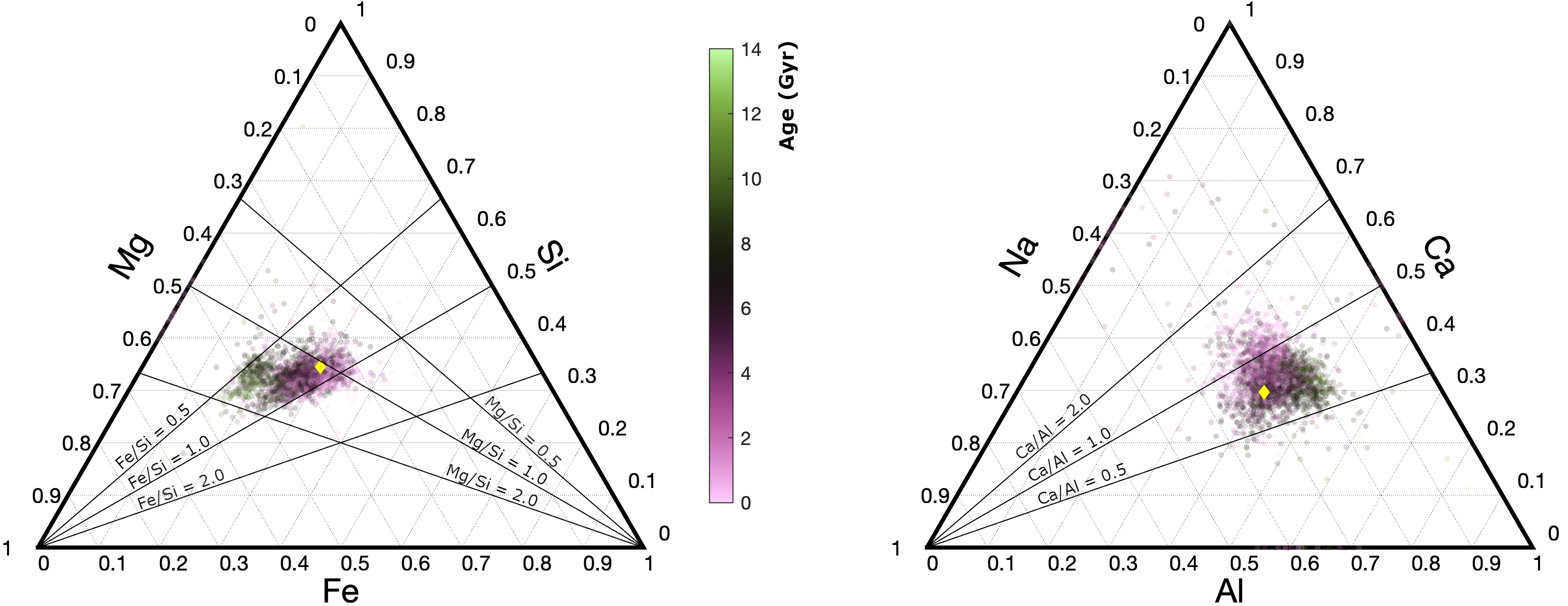}}
      \caption{Stellar abundances as contained by the Hypatia catalogue \citep{Hypatia}, colour-coded for stellar age as estimated from stellar Y/Mg and Y/Al (see equations 6 and 7 from \cite{Spina2018}). Solar composition from \cite{Lodders2009} is plotted for reference.}
      \label{fig:Suppl_Agetriangles}
\end{figure*}

\begin{figure*}[h]
    \resizebox{\hsize}{!}{\includegraphics{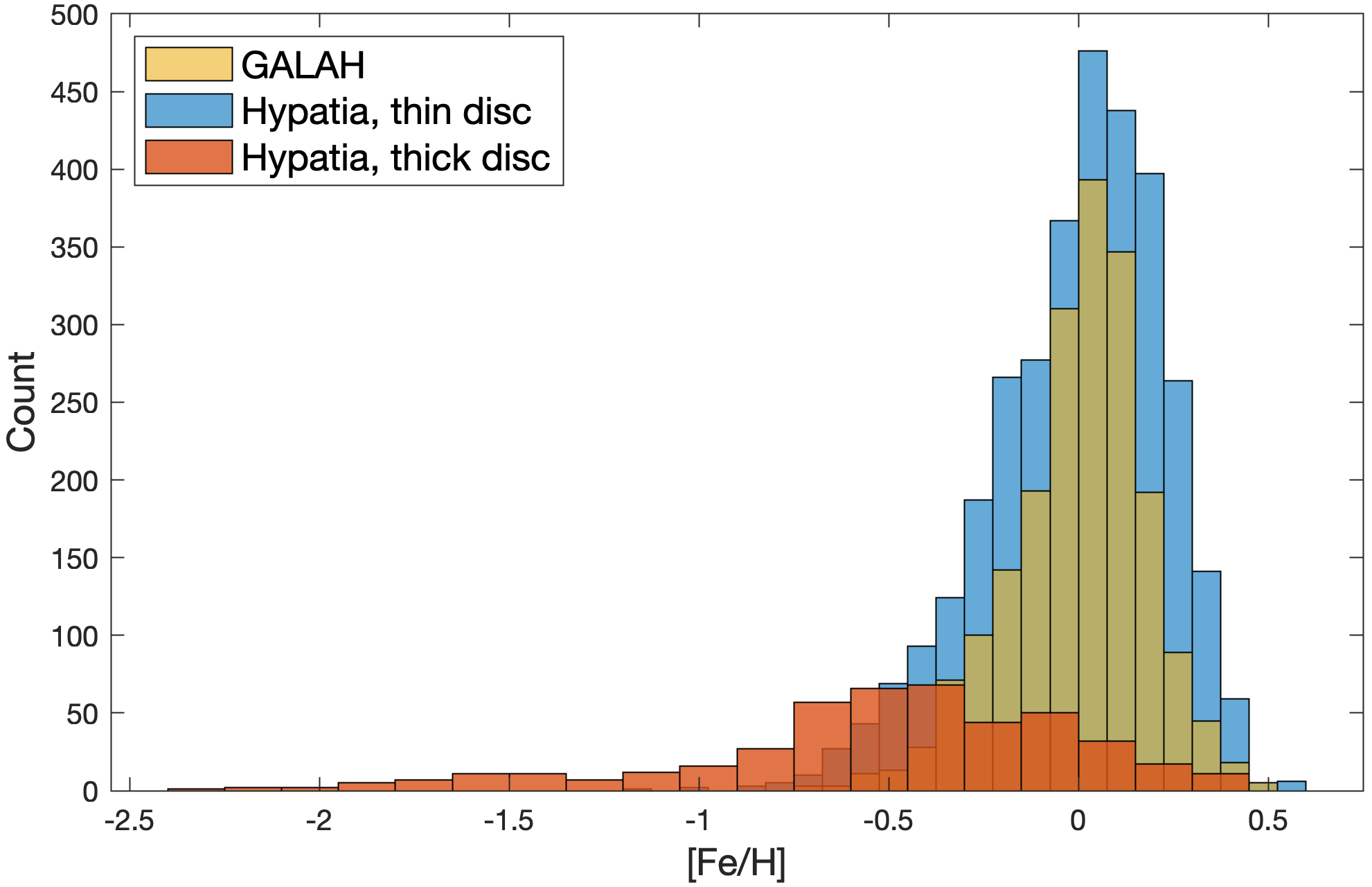}}
      \caption{Distribution of stellar metallicity (in terms of [Fe/H] in dex) of our sample. We distinguish between data from the GALAH catalogue (green), and the Hypatia catalogue for stars from the Galactic thick (brown) and thin (blue) discs.}
      \label{fig:Suppl_Fe_H}
\end{figure*}

\begin{figure*}[h]
    \resizebox{\hsize}{!}{\includegraphics{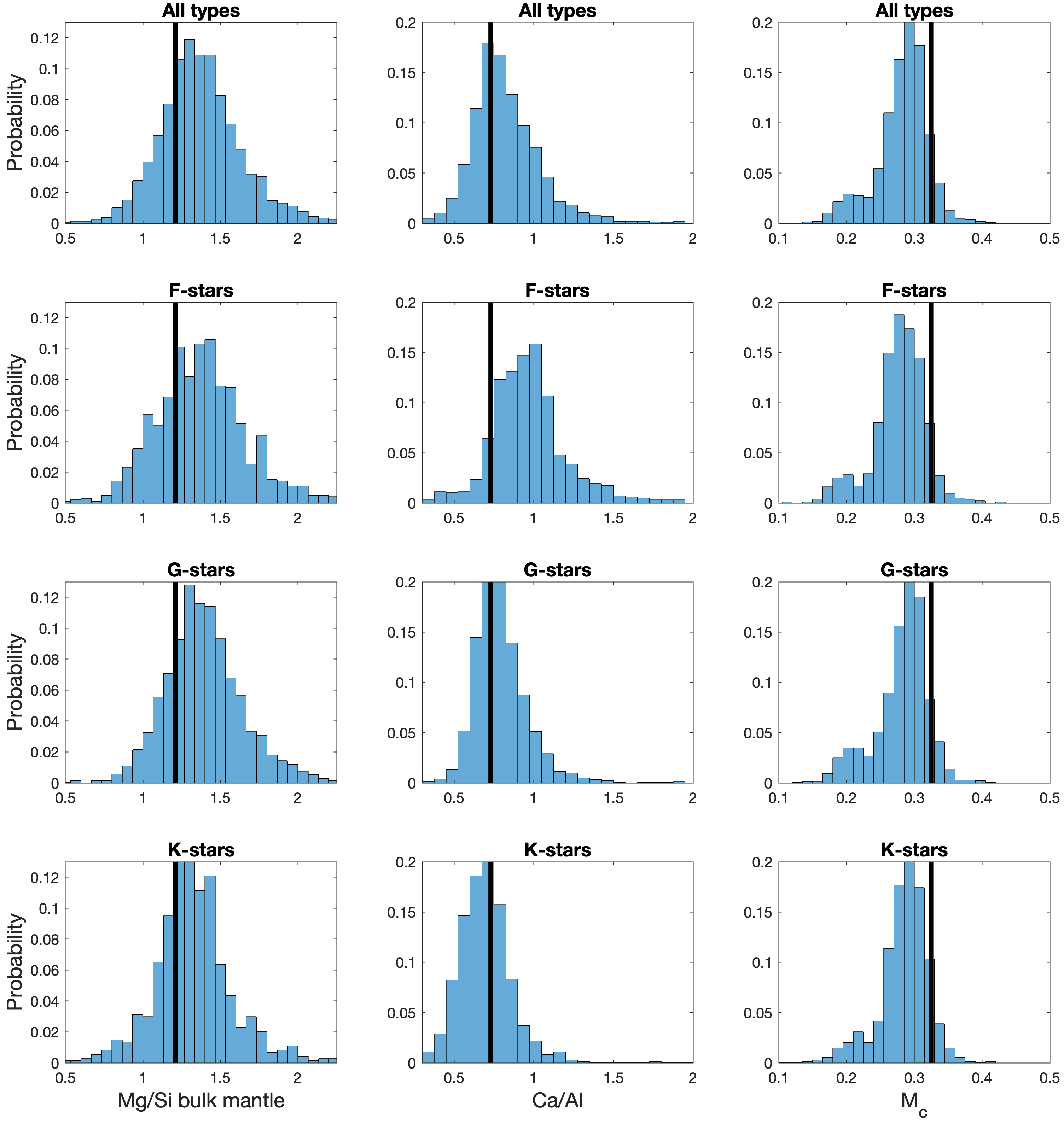}}
      \caption{Modelled planet compositions as a function of spectral class. Comparison of mantle molar Mg/Si (left), mantle molar Ca/Al (middle), and core sizes (right) of our population for all stars and stellar spectral types F, G, and K (top-to-bottom). Bulk Earth composition is plotted as a solid black line for comparison \citep{McDonough2003}.}
      \label{fig:Suppl_SpectralType}
\end{figure*}

\begin{figure*}[h]
    \resizebox{0.5\hsize}{!}{\includegraphics{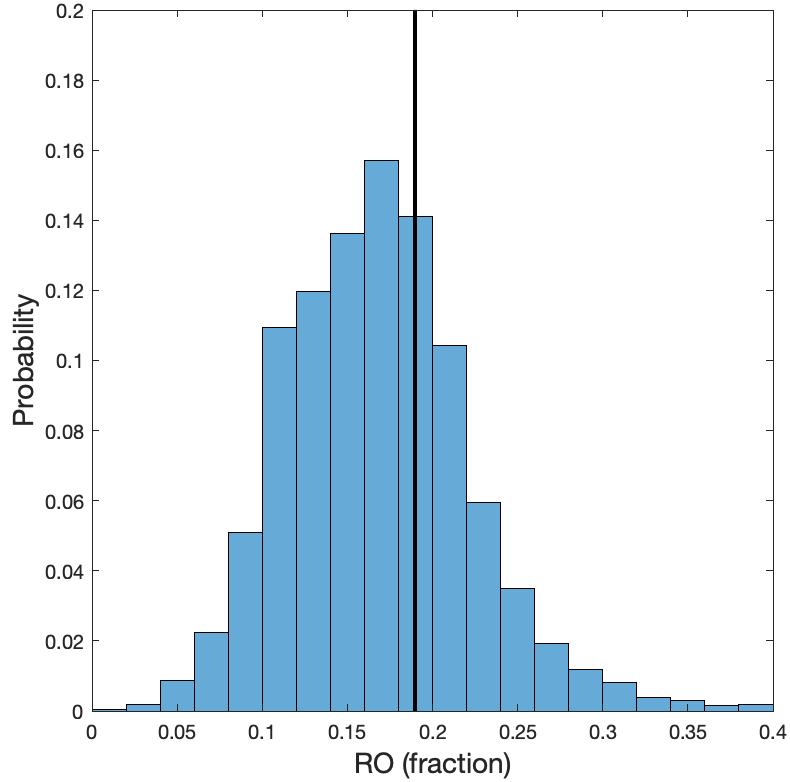}}
      \caption{Histogram of refractory oxygen fractions (RO) in our population under the assumption of constant \textit{fO$_2$}. This is the fraction of available oxygen that condensed as refractory compounds during planet formation, and is equivalent to the depletion factor for oxygen (see Tab.\ \ref{tab:depletion_factors}). The Earth-Sun RO fraction is plotted for comparison \citep{Wang2019}.}
      \label{fig:Suppl_RO}
\end{figure*}

\end{document}